PROGRESS SEMINAR

# 3D MPSoC with On-Chip Cache Support – Design and Exploitation

**RODRIGO CADORE CATALDO**

Advisor: César Augusto Missio Marcon

Co-Advisor: Debora da Silva Motta Matos

Porto Alegre, Brazil

July 2015

# LIST OF FIGURES

# LIST OF TABLES

# LIST OF ACRONYMS

| | |
|---|---|
| **2D** | Two-Dimensional |
| **3D** | Three-Dimensional |
| **API** | Application Programming Interface |
| **BEOL** | Back-End Of Line |
| **CABA** | Cycle Accurate Bit Accurate |
| **CCI** | Cache Coherent Interconnect |
| **CMOS** | Complementary Metal-Oxide-Semiconductor |
| **CMP** | Chip-MultiProcessor |
| **D2D** | Die-to-Die |
| **D2W** | Die-to-Wafer |
| **DMA** | Direct Memory Access |
| **DRAM** | Dynamic Random Access Memory |
| **DWM** | Domain Wall Memory |
| **FEOL** | Front-End Of Line |
| **FIMS** | Field Induced Magnetic Switching |
| **FS** | Full System |
| **GIC** | Generic Interrupt Controller |
| **GNU** | GNU's Not Unix |
| **GPL** | General Purpose License |
| **HMC** | Hybrid Memory Cube |
| **IC** | Integrated Circuit |
| **IDE** | Integrated Development Environment |
| **IPC** | Instruction Per Cycle |
| **ISA** | Instruction Set Architecture |
| **ISS** | Instruction Set Simulator |
| **LPDDR** | Low Power Double Data Rate |
| **LRU** | Least Recently Used |
| **MPP** | Massively Parallel Processor |
| **MPSoC** | Multiprocessor System-on-Chip |
| **MRAM** | Magnetoresistive Random Access Memory |

**MTJ**       Magnetic Tunnel Junction
**NMOS**      N-type Metal-Oxide-Semiconductor
**NUCA**      Non-Uniform Cache Architecture
**NUMA**      Non-Uniform Memory Access
**NoC**       Network-on-Chip
**NORMA**     NO Remote Memory Access
**O3**        Out-of-Order
**OS**        Operating System
**OSCI**      Open SystemC Initiative
**OVP**       Open Virtual Platforms
**PCRAM**     Phase-Change Random Access Memory
**PE**        Processing Element
**PIPT**      Physically Index Physically Tagged
**PPGCC**     Programa de Pós-Graduação em Ciência da Computação
**RTL**       Register Transfer Level
**SCU**       Snoop Control Unit
**SE**        System-call Emulation
**SMP**       Symmetric MultiProcessing
**SoC**       System-on-Chip
**SRAM**      Static Random Access Memory
**SSE**       Streaming SIMD Extensions
**STT**       Spin-Transfer Torque
**TAS**       Thermally Assisted Switching
**TLM**       Transactional Level Modeling
**TLM-DT**    TLM with Discrete Time
**TSV**       Through Silicon Via
**UART**      Universal Asynchronous Receiver/Transmitter
**UMA**       Uniform Memory Access
**VFP**       Vector Floating-Point
**VIPT**      Virtually Index Physically Tagged
**VNC**       Virtual Network Computing
**W2W**       Wafer-to-Wafer

# CONTENTS

# 1  INTRODUCTION

The evolution of technologies employed to manufacture Integrated Circuits (ICs) and the ever-increasing ratio of transistors per unit area brought the possibility of integrating billions of transistors in a single chip. This allowed the creation of the complete system functionality on a single IC called System-on-Chip (SoC). [INT15a][QUA15] are examples of commercial SoCs.

Following such progress, the demand of functionality and responsiveness of chips has also grown, boosting the development of the Multiprocessor System-on-Chip (MPSoC). This architecture is composed of multiple and possibly heterogeneous Processing Elements (PEs), a memory hierarchy (i.e., cache layers and main memory) and I/O components. Future MPSoCs will be made up of hundreds of such PEs [FER12][ITR11]. However, the growth of the quantity of PEs into a single chip requires a scalable interconnection to maintain acceptable communication latency and throughput. The traditional solutions based on buses can only handle few PEs and cannot scale to higher degrees of parallelism [BEN02]. Two-dimensional (2D) Network-on-chip (NoC) has emerged as a communication architecture that can overcome this problem through a scalable, packet-based network, inspired by years of study in computer networks [AHM10][JIA11]. To improve data communication and throughput, chip designers proposed three-dimensional (3D) NoCs, where components are distributed through stacked layers, instead of linearly in a single layer [AHM10][FRE12]. Consequently, reducing some complexity involved in the fabrication of ICs, such as global wires length [FEE09][FIC13]. Figure 1 shows an example of reduced wire length due to stacking of layers. In addition, 3D topologies may reduce the hop count when compared to their 2D counterpart [MAR14][MAR14b]. According to Feero et al. [FEE09], there are 40% more hops in a 2D mesh compared to that in a 3D mesh. Conversely, heat dissipation is exacerbated by having multiple layers of IC.

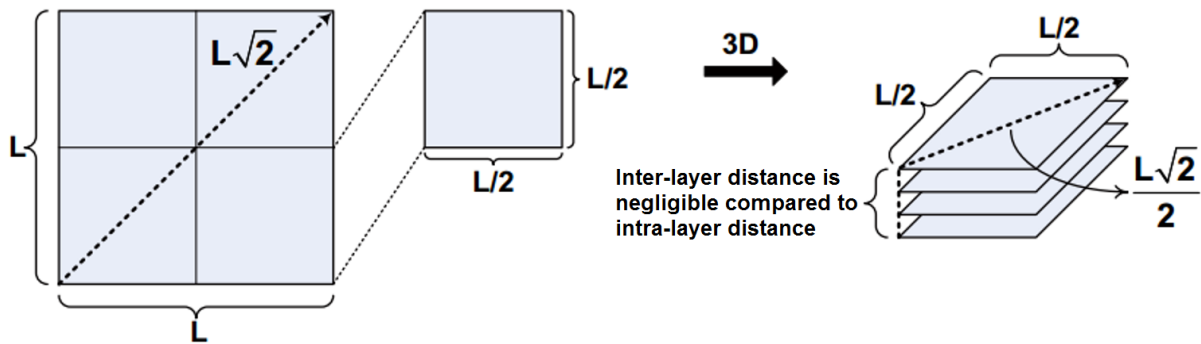

**Figure 1. Wiring length of simple layer vs a stack of smaller layers [LI06].**

Unfortunately, the use of packet-based transmission as the main network for communication compromises the efficiency of the memory hierarchy. A L1 cache area can be easily absorbed by the core area, but no so easily in the L2 cache case. Coskun et al. [COS09] demonstrated that, in 90 nm technology, an UltraSPARC T1 Core occupies 10 mm² of area and a L2 shared cache consumes 19 mm² of area. This means that the impact of cache architectures must be consider in the chip area. Some approaches for this situation replace some of the PEs in the NoC with shared L2 caches. Therefore, any access to a L2 cache must traverse the NoC twice: one for requesting the content of a given address and another one for the reply. As shown by previous works [FU14][WA08][YE10], even conservative rates of packets injection (i.e., ≤ 1%) result in large 2D NoC latencies. For instance, Ye et al. [YE10] shows that two L2 cache blocks distributed into an 8×8 NoC and packets injection rate of only 0.2% results in 1976 cycles of network latency, in average – a prohibitive latency for such small injection rate. These works have also demonstrated that on-chip traffic congestion is primarily caused by the intensive memory access requests and responses. Thus, a better design must be employed to efficiently tackle this problem [YE10].

## 1.1 Motivation

3D ICs present a promising technology for cache architecture schemes in a NoC based design. This technology enables the efficient manufacture of both logic and memory into a single IC, as they can be manufactured independently and integrated in a latter process [LOI10]. In addition, such fabrication process enables the use of emerging new technologies, such as PCRAM (Phase-Charge Random Access Memory) and MRAM (Magnetoresistive Random Access Memory). Both of these memory technologies are non-volatile, do not use basic CMOS (*Complementary Metal-Oxide-Semiconductor*) nor NMOS (*N-type Metal-Oxide-Semiconductor*) technologies and present much higher degree of density than SRAM (Static Random Access Memory), which is commonly used as the memory technology for caches. Secondly, 3D NoCs have emerged to reduce the length and the number of global interconnections that packets must pass through, and consequently, enabling to decrease the network latency and to increase its throughput. Nonetheless, the latency and throughput requirements for memory systems are hardly fulfilled by NoC architectures, since memory access rate are normally orders of magnitude higher than message exchange rate, even for IO bounded systems. Aiming to overcome these scenarios, this work proposes a two abstract layer system that employs disparate communication architectures for the inter-processor and the memory system communication; i.e., the inter-processor communication is performed by a NoC, whereas a special purpose architecture fulfills the memory communication requirements.

## 1.2 Contribution

The main contribution of this work is the design and experimental exploration of 3D MPSoCs with on-chip cache support that employs independent infrastructures for inter-processor and memory system communication. We propose the use of packet-based NoC for inter-processor communication for its efficiency of travelling small packets and its benefits to ever increasing scalability requirements [BEN02]. For the memory system, we propose the use of a cache coherence hierarchy implemented in a crossbar-based infrastructure. However, as coherence is costly to maintain and does not scale efficiently [MAT10], our system has a two-layer architecture. The first layer is the interconnection among PEs of a cluster presenting a single coherent address space and a Uniform Memory Access (UMA) model. The second layer interconnects clusters through a NO Remote Memory Access (NORMA) model – i.e., clusters do not have a shared address space. As such, communication between clusters are accomplished through a NoC architecture, which has the potential to scale to hundreds of PEs. Additionally, aggregating UMA and NORMA models in the same target architecture enables us to use multiprocessing and multicomputer programming jointly, which enlarges the exploration and implementation spectrum of highly-complex applications. Finally, experiments are being performed through the Gem5 simulator [BIN11] due to its ability to accurately model a coherent memory system, while being several times faster than a hardware-level model. The target architecture is based on the ARM Versatile Express development board [ARM15a].

## 1.3 Objectives

This work proposes the design of 3D NoC-Based MPSoCs with stacked memory layers, which is satisfied with the following strategic and specific objectives.

Strategic:

1. To explore available simulation tools for memory hierarchy evaluation;

2. To understand and to explore cache design for on-chip shared use;

3. To explore 3D MPSoC architectures taking into account a memory centric design (i.e., the fulfillment of the memory system requirements);

4. To analyze features of emerging/promising memory technologies (e.g., read/write access time, area consumption, read/write energy consumption, write endurance);

5. To select appropriate memory technology (i.e., taking into account memory features) for each level of the memory hierarchy; and

6. To assess the impact of memory technologies on MPSoC performance.

Specific:

1. To implement and to validate 3D MPSoCs with shared L2/L3 caches for latency and throughput evaluations. These evaluations are achieved through a benchmark suite, aiming to substantiate that such design is efficient to fulfill the ever-increasing demand of MPSoCs;

2. To explore emerging/promising memory technologies that have distinct characteristics from the commonly used DRAM and SRAM technologies.

## 1.4 Organization

This work is organized as follows. Section 2 discusses cache organization commonly found in MPSoC and multicore architectures and details the major memory architecture designs. Section 3 discusses and summarizes the basic design principles in developing current and emerging memory technologies. Section 4 presents related work of 3D MPSoC architectures with cache support. Section 5 presents the Gem5 simulator – its features, accuracy and shortcomings – and an overview of alternative modern full system simulators. Section 6 discusses the design and experimental exploration of 3D MPSoC in this work. Finally, Section 7 shows the schedule for this work development.

## 2 CACHE HIERARCHY IN MPSOC ARCHITECTURES

The memory hierarchy in the MPSoC architecture follows the same basic characteristics than the hierarchy present in the multicore architecture [MAR05b][ASA09]. The levels in it represent the tradeoff between latency and cost. The hierarchy can be summarized from the highest level of performance to the lowest as follows: multilevel caches, main memory and massive storage. The lowest levels of this hierarchy, main memory and massive storage, are frequently placed outside of manycore/multicore chips since they do not have a suitable response time when compared to the CPU and require large amounts of chip area. Nonetheless, the highest levels are frequently placed inside the chip.

For MPSoC architectures, the first level (i.e., L1 cache) of a multilevel cache hierarchy can be easily integrated into the processor area, because they have few kilobytes of memory. It does not happen with lower levels of cache due to their bigger size.

Asaduzzaman, Sibai and Rani [ASA09] classify the cache architecture in *distributed* and *shared* L2 cache, which are described next.

*The distributed L2 cache* where each core has a dedicated level-2 cache. Figure 2 shows a schematic representation of an MPSoC with this organization. The L1 cache area is divided between instruction cache (I1) and data cache (D1). The communication architecture is used iff a memory request misses in both L1 and L2 cache banks.

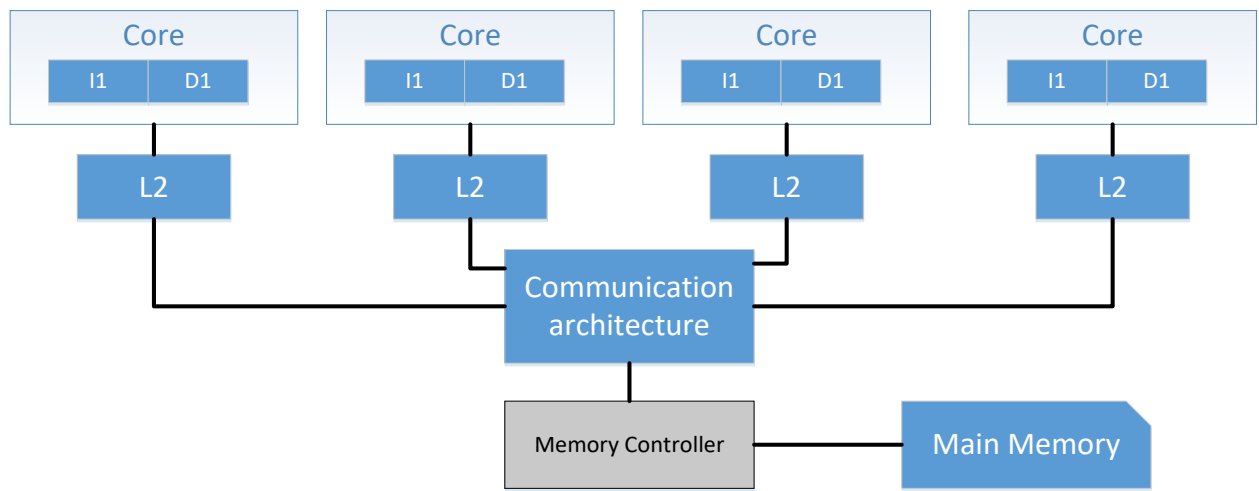

**Figure 2. Schematic representation of an MPSoC with distributed L2 cache blocks (based on [ASA09]).**

An actual product from Intel called Xeon Phi uses a similar architecture and it is shown in Figure 3. Each core is equipped with a 32KB L1 instruction and 32KB L1 data cache, and an individual 512 KB L2 cache bank. The entire cache system is kept coherent. This product uses a bidirectional ring for communication. Each direction is comprised of three independent rings for data, address and acknowledgment messages [INT15b]. The next generation of the Xeon Phi, called Knight's Landing, is an upcoming chip that features stacked memory chips linked through TSV (*Through Silicon Via*) to greatly increase the amount of memory bandwidth that can be feed to the cores [IDG15]. Full details of this platform is still not available, but it is expected to contain at least 60 cores [PLA15][STO15].

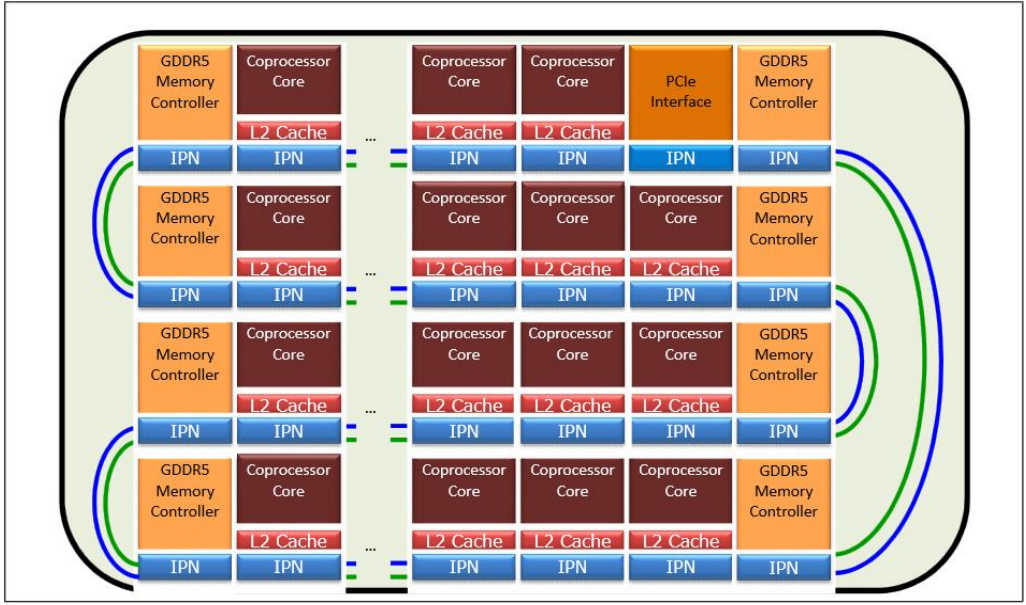

**Figure 3. Conceptual diagram of the structure of the Intel Xeon Phi coprocessor [INT14].**

Tilera Corporation is a semiconductor company focusing on scalable multicore embedded processor design. Its products range from supporting 4 to 200+ cores [TIL11a]. In the Tilera-Gx architecture, each core has 32KB L1 instruction and 32KB L1 data cache, and an individual 256KB L2 cache. Figure 4 depicts the core unit, memory subsystem and the communication network. Tilera-Gx uses five independent NoCs, whereas three of them are related directly to the memory subsystem. The mesh topology is employed in this product. The rationale for independent networks is to allow low latency communication for a scalable architecture [TIL11c].

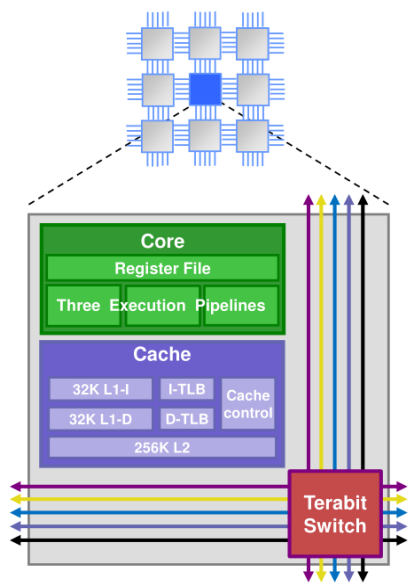

**Figure 4. General purpose core unit of Tile-Gx100 [TIL11b].**

IBM produces Power8 processors that have 12 cores per chip. Every core has access to 64KB of L1 data and instruction caches. In addition, each core has an individual 512KB SRAM L2 cache. They share a 96MB eDRAM L3 cache and an optional eDRAM L4 cache. Chip interconnect is accomplished through buses [IBM14][STU13]. From L2 to lower levels, a NUCA (*Non Uniform Cache Architecture*) is used – which means that not all accesses to L2 have the same latency. This architecture will be discussed later.

The second classification is denominated *shared L2 cache*, where, one or more L2 blocks are distributed across the chip to service the MPSoC architecture. Figure 5 depicts a schematic representation of an MPSoC with this organization. As the case of Figure 2, the L1 cache area is divided between instruction cache (I1) and data cache (D1). The NoC is used iff a memory request misses in the L1 cache bank.

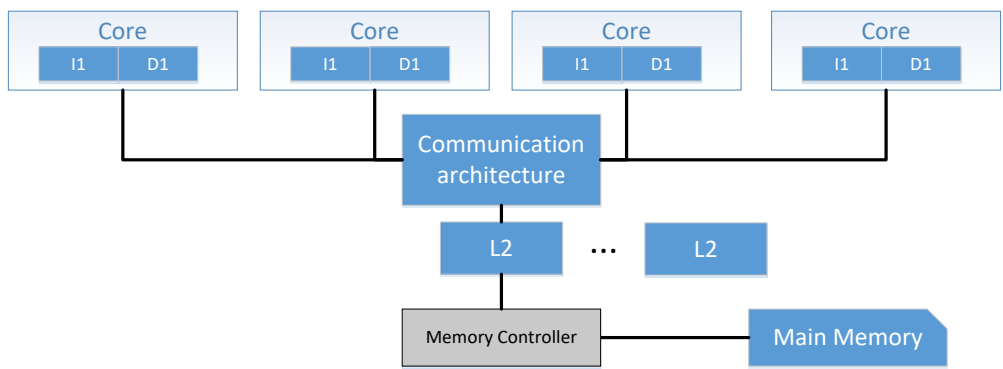

**Figure 5. Schematic illustration of an MPSoC with shared L2 caches (based on [ASA09]).**

Figure 6 shows a product from Fujitsu with a similar architecture called SPARC64-X. This product has four shared L2 caches of 6MB each (up to 24MB).

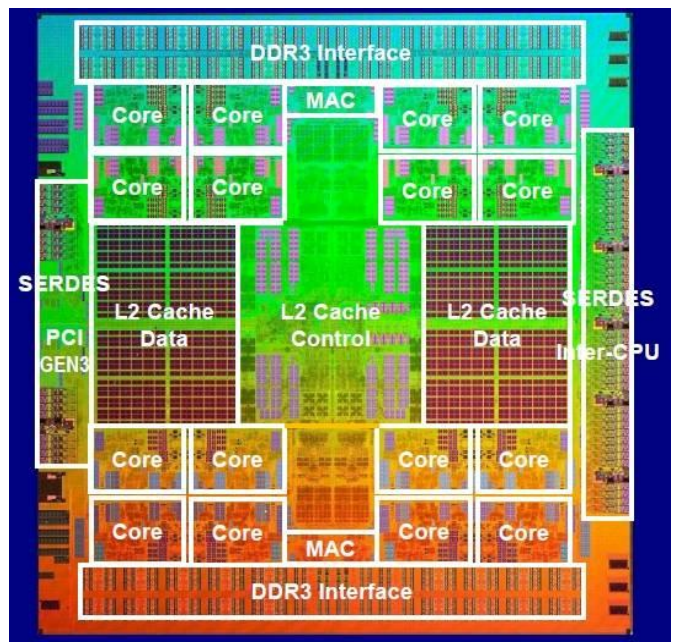

**Figure 6. SPARC64-X die: sixteen cores and four banks of L2 cache [REG15].**

Oracle proposed in 2014 a 32-core chip called SPARC M7. The cores are organized in 8-core clusters, where each core accesses 16KB of instruction and data cache. Pairs of two cores have access to a shared 256KB L2 cache of data. Finally, each pair of four cores has access to a 256KB L2 cache of instructions and a L3 cache of 8Mb [LI15][SIV14]. Figure 7 shows the cache hierarchy and the chip organization of SPARC M7.

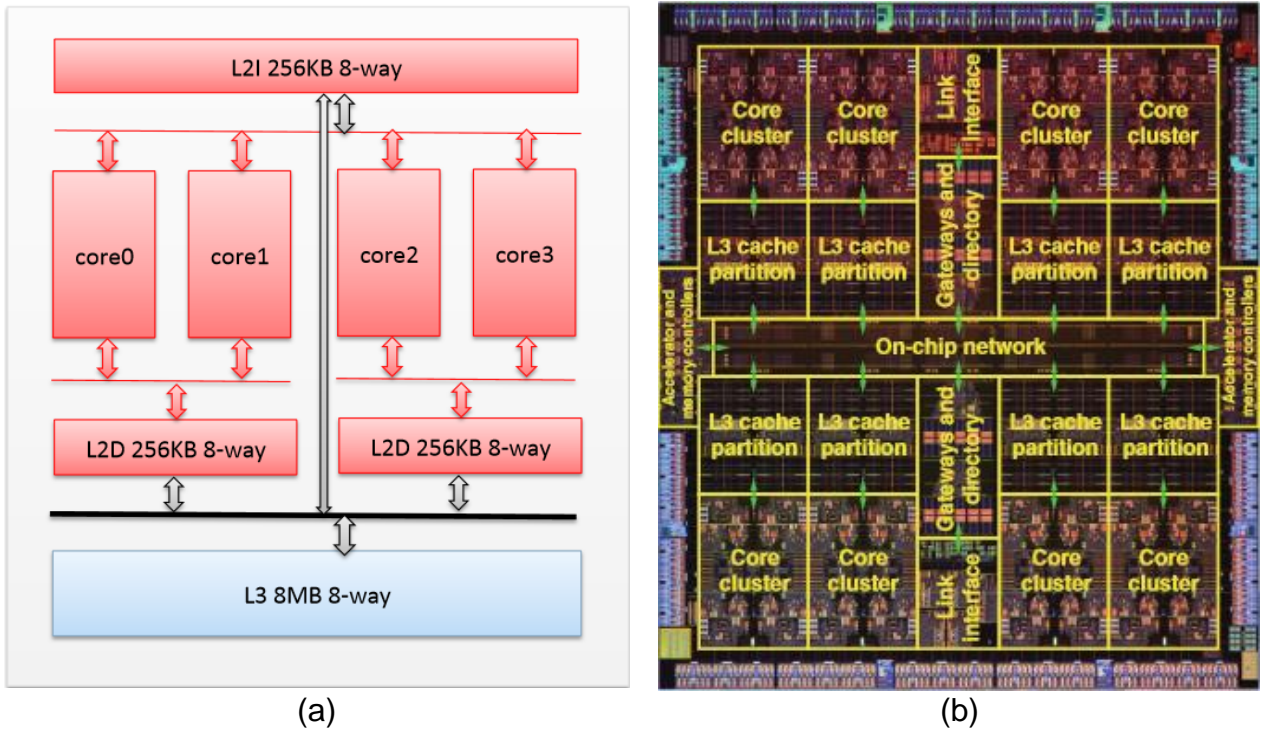

(a)                                    (b)

**Figure 7. SPARC M7: (a) schematic of core cluster design, (b) chip layout [SIV14].**

The crossbar-based network used in previous SPARC processors is scrapped, and a new mixed network is employed. This network connects all L3 caches and four memory controllers. The network is logically made of three physical networks: a request network with a 4-ring topology (maximum hop count is 11), a point-to-point response network and a multistage mesh of six 10×10 switches [AIN15][LI15]. Figure 8 depicts the organization of two of the three physical NoCs – the omitted request network overlies physically the data network.

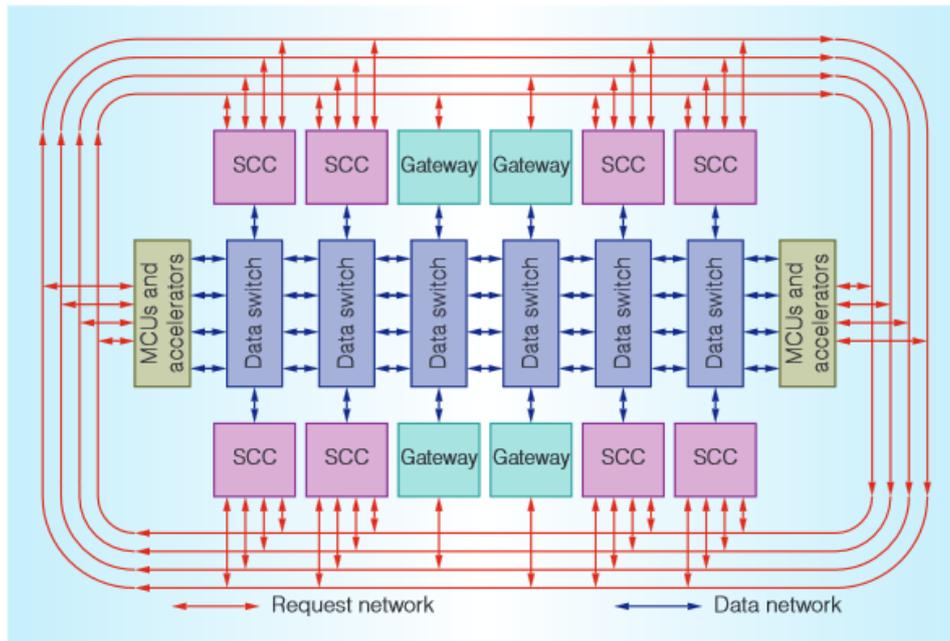

**Figure 8. On-chip network request and data topologies. The response network is omitted in this representation [AIN15].**

ARM has proposed an architecture called big.LITTLE that pairs a high-performance out-of-order processor cluster with a low-power in-order processor cluster to deliver optimal performance and energy consumption [ARM13a]. The two processor clusters are architecturally compatible, meaning that they can exchange tasks. Therefore, the high-performance processor cluster can shut down, when no application requires such performance and only the more modest processor cluster can handle the system. Any time that performance requirement is increased, the high-performance processor cluster is initialized and can 'steal' tasks from the modest cluster [ARM13b]. Each cluster has its own L2 cache. Figure 9 shows the Exynos 5 Octa SoC – one of the products that uses the big.LITTLE architecture. For this SoC, the L2 cache has 2MB and 512KB of space for, respectively, the high-performance cluster and the low-power processor cluster [SAM15b]. Communication is accomplished through a multilayer bus.

Both, distributed and shared L2 cache architectures present tradeoffs of performance, area and energy consumption [ASA09]. Distributed L2 cache alleviates traffic congestion in the interconnect fabric, since each core has its own L2 cache. However, this architecture has the initial cost of feeding data to so many caches. In addition, the area consumption of L2 cache is not negligible [COS09]. In juxtaposition, shared L2 cache increases traffic congestion in the interconnect fabric since each memory request for an L2 cache must traverse it. The initial cost of feeding data for a shared L2 cache can be mitigated if more than one core shares the same code data. In addition, shared L2 caches are better adjusted to area restriction due to their reduced quantity when compared to the previous architecture. Asaduzzaman et al. [ASA09] show experimental results of both architectures in multicore systems. They conclude that the impact of adding cores to the system is in favor of the distributed architecture. However, the three applications used for testing are limited to the same characteristics: multimedia application profile and low code size that fits entirely in the L2 cache (1 to 2.5KB).

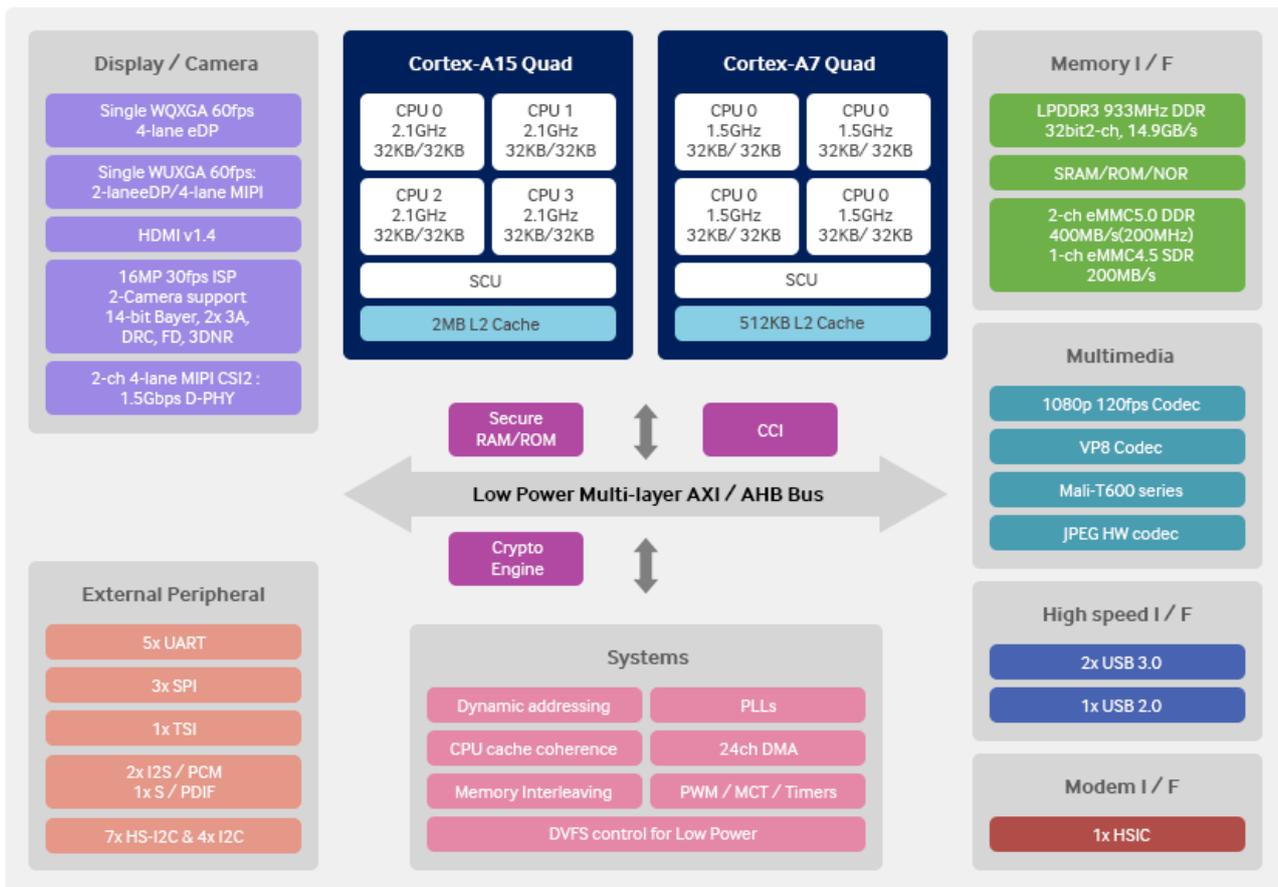

**Figure 9. Architecture of the Exynos 5 Octa SoC [SAM15b].**

## 2.1  Memory Architecture Design

Given the inherent difficulty of writing programs to run well in parallel systems, one feature often found is the ability to address the whole physical memory space as a single entity. Thus, the programmer does not need to concern the data placement, because all variables are accessible at any time to any processor [PAT13]. This type of system is named Symmetric MultiProcessing (SMP). When the physical address is a unique entity, the hardware typically provides cache coherence to give a consistent view of the memory subsystem [MAR12][PAT13]. The two classifications discussed in the previous subsection uses such coherent view.

For single address space, two of the most common architectures are UMA and NUMA (*Non-Uniform Memory Access*) [PAT13]. In the first architecture, all processors access any memory position with the same latency. In the second architecture, processors access the same memory position with different latencies.

Performance in the UMA architecture is limited to the communication capacity between processors and memories. Adding processors to the system beyond some point does not increase performance linearly, since they share the same memory bandwidth [GEN12]. Thus, scaling beyond the dozens of processors requires a NUMA architecture [HWA12]. The non-uniformity of access can also be applied to an individual memory unit, whereas the access latency depends on the location requested. Traditional L2 caches were comprised of a single bank, which had an average high latency access to protect the worst-case scenario. Now, L2 caches are divided in blocks to allow parallel access and diminish the individual bank latency [ARM11a][OLU07]. However, such design increases the latency penalty for the worst-case scenario as shown in Figure 10.

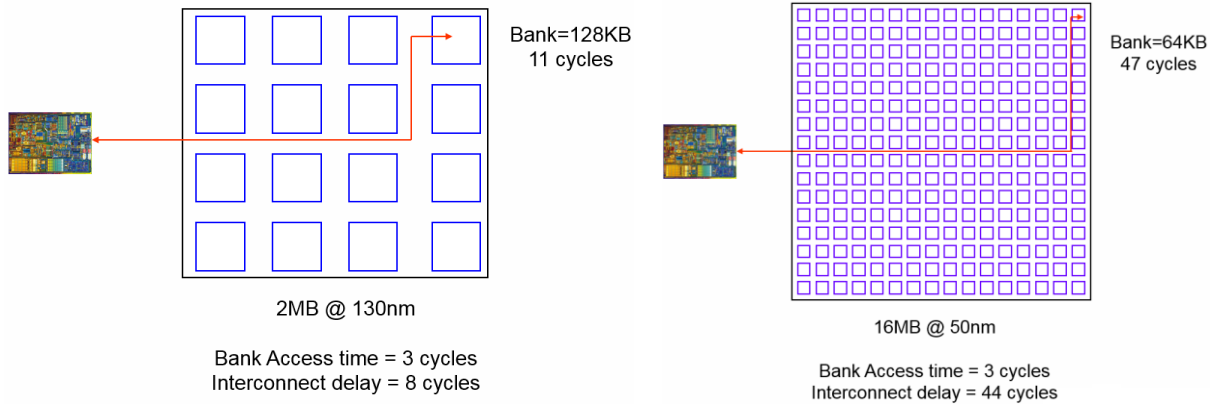

**Figure 10. Effect of increasing data banks on traditional L2 caches [LEE15].**

Aiming to minimize this penalty, the NUCA architecture was proposed by Kim et al. [KIM02]. The high latency access due to worst-case scenario can be avoided if the cache controller can access each bank at a speed proportional to its distance from the cache controller. This opens new design space for exploration of cache policies regarding mapping, search algorithm and data migration [KIM03]. In addition, exploiting the disparate latencies, it is possible to design cache memories that have more than one technology embedded in the same unit. Wu et al. [WU09] study the effect of using disparate memory technologies in both intra and inter cache levels. The intra cache level is when a single cache level is partitioned into multiple regions, where each region exploits the advantage of the memory technology employed (i.e., higher density, lower power dissipation, etc.). The inter level cache has the same principle but uses multiple levels of cache to employ more than one technology. Experimental results with a full system simulator showed that the intra level cache can provide 12% of IPC improvement over a 3-level SRAM cache design under the same area constraint, in average, and; Inter level cache can provide 7% of IPC improvement on the same conditions. Section 3 will discuss current and emerging memory technologies.

Future MPSoCs, made of hundreds of processing units [FER12][ITR11], hinder the ability of the hardware to provide a coherent view of the entire memory space as proposed by the UMA and NUMA architecture. For large scale systems, NORMA is attractive due to its ability to decentralize resources and increase reliability [HWA12]. Processor communication is carried out by message passing through the NoC [HWA11], and through network protocols, more than one operating system can be easily integrated.

## 2.2 Cache Organization in 2D MPSoC Architectures

For 2D MPSoC L1 cache is predominantly integrated into the core area. Both architectures of L2 caches are possible; however, there is not still a dominant architecture paradigm for L2 caches [SAB10]. Figure 11 shows two of eight L2 cache studied by Ye et al. [YE10]. This study clearly shown that the dominant factor of L2 access latency was the packet-based NoC latency. Even conservative injection rates of packets (≤ 1%) resulted in expressive network congestion on mesh topologies such as 8×8. In addition, this study and the study done by Wang et al. [WA08] conclude that on-chip traffic congestion is predominantly caused by the intensive memory access of requests and responses.

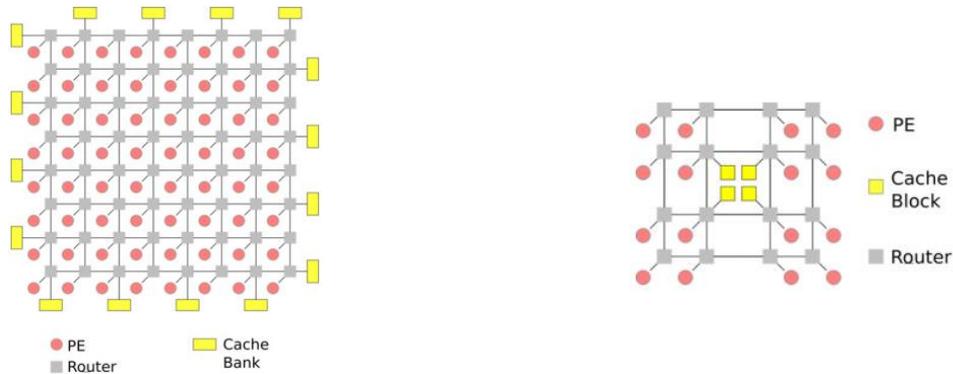

**Figure 11. Two of eight L2 cache configurations studied on a 2D MPSoC [YE10].**

Cho et al. [CHO06] present a mixture of shared and distributed L2 cache architectures. They argue that none of them achieves optimal performance under diverse workloads. Hence, in this work, each core has a L2 slice controlled by the operating system (OS). A L2 slice is a smaller set of a full L2 cache. The OS controls where a cache line will be placed, locally or remotely, resulting in different access times. This means that the OS can choose to use any cache architecture according to its policies. Figure 12 shows the cache organization in this work.

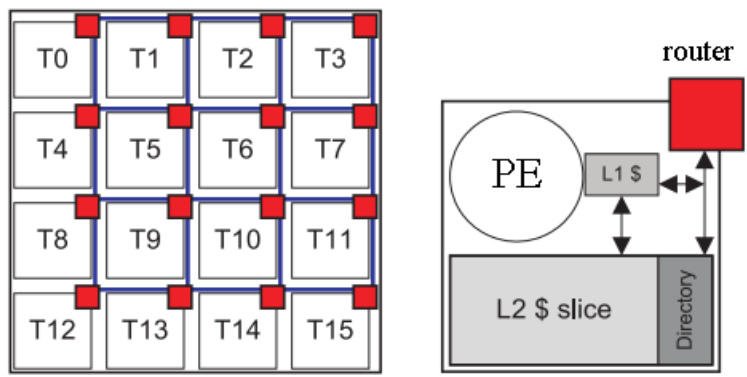

**Figure 12. Sixteen core tiles (left) and the design of a single core (right) (based on [CHO06]).**

## 2.3  Cache Organization in 3D MPSoC Architectures

3D MPSoCs introduced potential new architectures that are not commonly found in their 2D counterpart. The exploration of such architectures was enabled by the ability to stack multiple dies with diverse fabrication processes. Examples of such explorations are the use of emerging memory technologies in lower cache levels and even the presence of main memory in stacked dies. These emerging technologies present much higher density when compared to traditional technologies with the cost of higher access latency. Section 4 summarizes recent work on 3D MPSoC architectures. Next section reviews the Through Silicon Via (TSV), which is the most employed technology for connecting multiple dies.

### 2.3.1 THROUGH SILICON VIA (TSV)

TSVs are basic building elements providing connections between different stacked dies. A TSV is a galvanic connection between two wafers [SII12]. Its manufacture consists of making a hole in the silicon, which is filled in another process step. Figure 13 shows an illustration of multiple 2D ICs that are thinned, bounded together and interconnected with TSVs distributed within the planes of the 2D ICs.

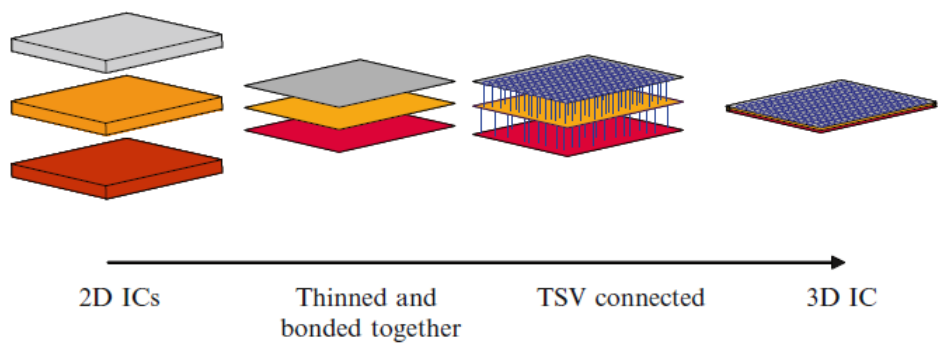

**Figure 13. Basic structure of a 3D IC [PAP11].**

Technological challenges are presented in this process since TSVs add considerable costs in 3D fabrication. Following, a short overview of these challenges are presented.

#### *Compound Yield*

The wide adoption of 3D ICs can only be fully performed when the design and manufacturing cost of these products are commercially viable. Among the various factors that affect such cost, manufacturing yield is one of the most crucial [SMI07]. A prototype of a three-layer chip in 2009 yielded up to 60% [MIY09] – however, recent researchers have reached better yield results [KIM13]. There are essentially two kinds of yield losses in 3D IC [XU12]: stack yield loss and assembly yield loss. Stack yield losses occur when one or more of the stacked dies have defects. Assembly yield losses occur by defects on the assembling process. The assembly process for 3D IC involves many challenging manufacturing steps, such as wafer thinning and wafer alignment. The method of 3D bonding chosen also has an impact on yield performance [SMI07].

#### *Method of 3D Bonding*

There are mainly three techniques used to stack ICs [SII12]: Wafer-to-Wafer (W2W), Die-to-Wafer (D2W) and Die-to-Die (D2D). W2W strategy stacks dies based on wafer alignment. D2D strategy extracts all dies first and then align them up. Finally, D2W strategy aligns die to wafers.

These techniques offer different trade-offs in production yield, production throughput and flexibility. In W2W technique, two wafers are stacked together and dies are extracted after assembling. This solution provides a high throughput but dies of different wafers must have the same dimensions [EVG15]. D2W and D2D do not have this limitation, because they allow testing dies individually before 3D assembly and, thus, offers a strong advantage in yield [SMI07]. In addition, W2W requires dies equal in size, while D2W and D2D allow dies with different sizes [SII12].

### TSV process steps

The main steps for the TSV fabrication consist of TSV drilling, TSV insulation, TSV metallization, Front End of Line (FEOL) formation, Back End of Line (BEOL) formation, handler attachment, wafer thinning and backside process.

**Table 1. Comparison of process flow for three processes of TSV manufacture [LI10].**

| Process Step\Processes | Via-First | Via-Middle | Via-Last |
|---|---|---|---|
| TSV Drilling | Phase 1 | Phase 1 | Phase 3 |
| TSV Isulation | Phase 1 | Phase 1 | Phase 3 |
| TSV Metallization | Phase 1 | Phase 2 | Phase 3 |
| FEOL Formation | Phase 2 | Phase 2 | Phase 1 |
| BEOL Formation | Phase 2 | Phase 2 | Phase 1 |
| Handler Attachment | Phase 3 | Phase 3 | Phase 2 |
| Wafer Thinning | Phase 3 | Phase 3 | Phase 2 |
| Backside Process | Phase 4 | Phase 4 | Phase 3 |

Table 1 presents the comparison of process flow for three different types of integration: Via-First, Via-Middle and Via-Last. For the Via-First process, the fabrication of TSVs is done before the Si FEOL device fabrication. The Via-Last process operates in the opposite direction: TSVs are fabricated after the Si FEOL formation. For the Via-Middle process, TSVs are fabricated after the Si FEOL, but before the Si BEOL formation.

Despite those technological challenges, foundries have already incorporated TSV and/or 3D ICs in their catalog. Examples of such cases are found in [CEA15][TEZ15a][ZIP15]. Additionally, Tezzaron [TEZ15b] presents a non-exhaustive list of organizations that are working with 3D technology.

TSV is at the center of one of the most important changes to the memory interface and, consequently, to the famous Memory Wall [WUL95]. The Memory Wall is the observation of the increasingly processor/memory performance gap in at least the last 20 years. On top of that, the trend of placing more and more cores on a single chip exacerbates this gap. One way to diminish this is to increase memory bandwidth through wider interfaces. However, until now this has been very challenging due to the fact that wider interfaces means more off chip pins and such pins are very expansive to add [FU14][ROG09][PAD11]. TSV eliminates such limitation by stacking dies and incorporating main memory into the chip, hence, no extra off chip pins are necessary.

Wide I/O is a standard to maximize the memory bandwidth at the lowest possible power dissipation. The key is to stack multiple memory channels on top of the system and interconnect them through TSV [CAD15a]. Recently, JEDEC[1] published the second standard of Wide I/O that presents significant improvements [JED15]. At half of the power dissipation of LPDDR (*Low Power Double Data Rate*) 3, Wide I/O can maintain the same memory bandwidth. Increasing the memory frequency, Wide I/O effectively provides more than double the baseline LPDDR bandwidth [GRE12][VIV11]. Samsung develops this technology since 2011 [SAM15c].

---

[1] JEDEC is a global leader in developing open standards for the microelectronics industry, with more than 3,000 volunteers representing nearly 300 member companies (source: //www.jedec.org/about-jedec).

Hybrid Memory Cube (HMC) is another memory solution that relies on TSV. While Wide I/O aims at the mobile low-power market, HMC aims at the high-performance server market. HMC achieves up to 15 times the bandwidth with 70% less energy consumption when compared to traditional DDR3 technology [CAD15b][MIC15]. In this memory, four to eight stacks of DRAM are on top of a single logic chip responsible for data access [HMC14]. The HMC consortium develops the HMC interface specification and promotes integration into a wide variety of systems [HMC15].

# 3   MEMORY SYSTEMS: CURRENT AND EMERGING TECHNOLOGIES

The dominant memory types in modern microelectronics devices are Dynamic Random Access Memory (DRAM), Static Random Access Memory (SRAM) and flash memory. For many decades, these memory technologies have been scaled down to achieve higher performance and to increase density at lower bit cost. However, these memories are gradually getting closer to their physical limits of scalability. Fortunately, new types of memory are being researched. These new types of memory present various degrees of tradeoff regarding scalability, operating voltages, power dissipation, retention time and operating speed [MAK12].

On a classic high performance embedded system, the memory hierarchy goes from the register and cache in high performance (normally implemented with SRAM), through the primary memory (normally implemented with DRAM) to a secondary memory (normally implemented with flash or electromechanical memory system for massive storage) [AER10][SAM12][SAM15a]. In this context, emerging memory solutions can open the way to new design architectures with full or partial replacement of such existing memories. The clear targets of emerging memories in the memory hierarchy are lowering energy consumption, increasing density and non-volatility of information.

3D ICs present an attractive design to build such systems with emerging technologies, since dies can be manufactured with heterogeneous technology [LOI10]. Stacking memories directly on top of an MPSoC is a natural way to attack the memory bottleneck, as stacking decreases the hop average in a NoC compared to a planar structure.

This section gives an overview on current and emerging memory architectures for 3D ICs, presenting some challenges and tradeoffs, and a conclusion comparing the memory technologies described here.

## 3.1  Static Random Access Memory (SRAM)

SRAM is a type of semiconductor memory that uses latches to store each bit. They do not need to be periodically refreshed (hence, static). Nevertheless, it is still a volatile memory since data is eventually lost when the memory is powered off. The basic structure of this memory uses six transistors to form a cell. However, other sizes are also available – ranging from four to ten [CHA00][YAM10]. Figure 14 shows the structure of a 6-transistor SRAM cell. They have two stable states, representing zero and one. The state is stable as long as power is available. Reading in this memory is done through the access line **WL**. After this signal is raised, the state of the cell is immediately available for reading on **BL** and $\overline{\textbf{BL}}$. If the cell state must be overwritten, the **BL** and $\overline{\textbf{BL}}$ lines are first set to the desired value and then **WL** is raised.

SRAM presents the following properties: (i) very fast access to read and to write data. Therefore, it is the preferred technology for caches in embedded systems; (ii) SRAM cell requires six transistors in the most basic design. Consequently, this technology has low memory density; and (iii) SRAM is more expensive in a cost/bit comparison when compared to the traditional DRAM memory [DRE07].

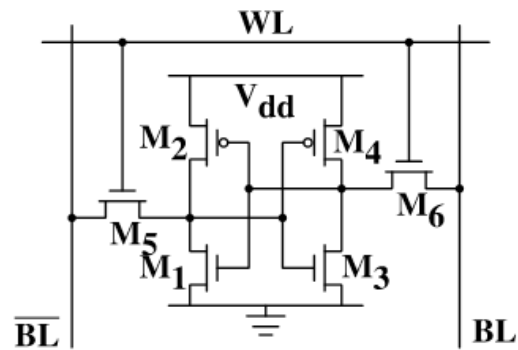

**Figure 14. Basic SRAM with six transistors [DRE07].**

## 3.2 Dynamic Random Access Memory (DRAM)

DRAM is a memory that stores each bit of data in a separate capacitor. Each capacitor can be charged or discharged; these two states are taken to represent one and zero. Figure 15 shows a DRAM cell with its capacitor **C** and transistor **M**. This transistor is used to guard the state access. For a read operation, the access line **AL** must be raised. This either causes a current to flow on the data line **DL** or not, depending on the charge in the capacitor **C**. For a write operation, **DL** is set to the desired value and **AL** is raised for a time long enough to charge or drain the capacitor [DRE07].

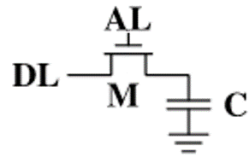

**Figure 15. A DRAM cell [DRE07].**

The logic information is stored in the capacitive load of the DRAM cell. In the course of time, a leakage current may discharge the DRAM's capacitor, causing loss of information. In order to avoid data loss, the data stored in the capacitor should be read and rewritten periodically. This process is called refreshing. During this period, no access to the memory is possible since a refresh is a read/write operation. Such operation waste energy and degrade performance by delaying memory requests [LIU12].

The DRAM presents the following properties: (i) fast access for read and write operations. Unlike SRAM, the output information is not immediately available, as it will always take some time until the capacitor discharges. Still, DRAM access time is considered fast when compared to other types of memories; (ii) DRAM cell is simpler and more regular than SRAM, which means packing many cells together in a die is simpler. Thus, DRAM presents more density and a better cost/bit tradeoff when compared to SRAM; and (iii) the periodic need for refreshing implies that DRAM is power-hungry [DRE07];

## 3.3 Embedded Dynamic Random Access Memory (eDRAM)

eDRAM has three commonly identified types of fabrication [TSM15]: DRAM-based, blended (or hybrid) and logic-based. DRAM-based technology is practically the same as commodity DRAM – a conventional one-transistor DRAM cell. Blended and logic-based technologies use additional mechanisms to enhance logic performance [TSM15]. eDRAM differs from DRAM in which the former is produced in a single chip that is part DRAM and part microprocessor, while the latter has processor and memory in separate packages. As such, eDRAM presents the following properties: (i) drastically reduced power dissipation when compared to DRAM; (ii) reduced manufacturing costs when compared to DRAM and SRAM; (iii) higher degrees of density when compared to SRAM; and (iv) a strong drawback - since eDRAM uses fast logic transistor, which have higher leakage than conventional DRAM, it needs shorter refresh intervals [MIT14][MON15][NEO01][TOS15].

## 3.4  Phase Change Random Access Memory (PCRAM)

PCRAM, also known as PCM, stores memory data in a chalcogenide, which is a phase-change material. Some common chalcogenide are alloy of germanium, antimony and tellurium [LEE10]. The memory cell resembles a capacitor like structure, where a group of alloys is sandwiched between two metal electrodes. Storage is done through the resistance change between a low resistance state (set operation) and a high resistance state (reset operation). Cell writing (i.e., set operation) is achieved through the operation of changing the cell from crystalline state to amorphous state. The opposite change, amorphous to crystalline state, is done for cell erasing (i.e., reset operation). Figure 16 shows the storage unit and the cell structure.

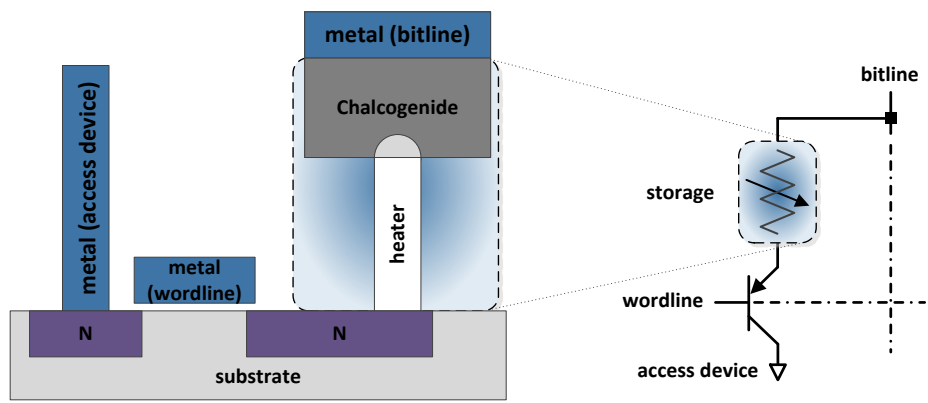

Figure 16. Storage element with heater and chalcogenide (based on [LEE10]).

PCRAM presents the following properties: (i) limited write endurance and high write latency, when compared to classic memory technologies such as DRAM. The write endurance of PCRAM is near of $10^8$, making this type of memory appropriated to use only in lower-level caches as L3 and L4, since they have less write traffic than higher-level caches (i.e., L1 and L2) [WU09]; (ii) it does not consume standby leakage power; and (iii) it is denser than DRAM [DON09a][MIT14].

## 3.5  Magnetoresistive Random Access Memory (MRAM)

MRAM is a non-volatile memory that uses magnetic elements for information storage. Figure 17 presents a schematic illustration of MRAM, whose basic element is a Magnetic Tunnel Junction (MTJ) that is sandwiched between two layers – a thin oxide barrier separated by two ferromagnetic layers. One of the two layers is a permanent magnetic body, which is set to a particular polarity (fixed due to fabrication process); in order to store new data, the field of the other layer can be changed to match an external field [MAK12][POR14].

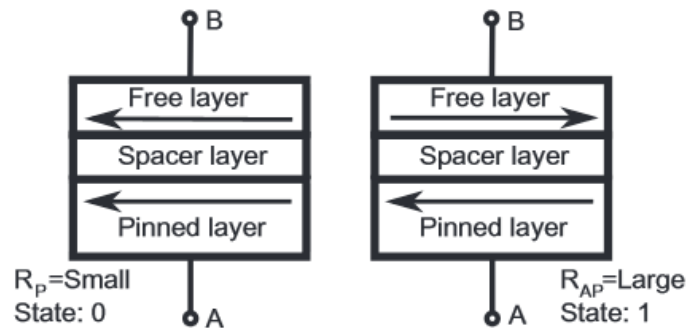

**Figure 17. Schematic of a 3-layer MTJ in low (left) and high (right) resistance state [MAK12].**

To read information, a connected transistor is turned on, and the memory state is determined by measuring the amount of current that flows through the bit line. Setting information is achieved by passing current through two perpendicular write lines, respectively termed "bit line" and "digit line" [AKE05]. Figure 18 shows the connected transistor and write lines.

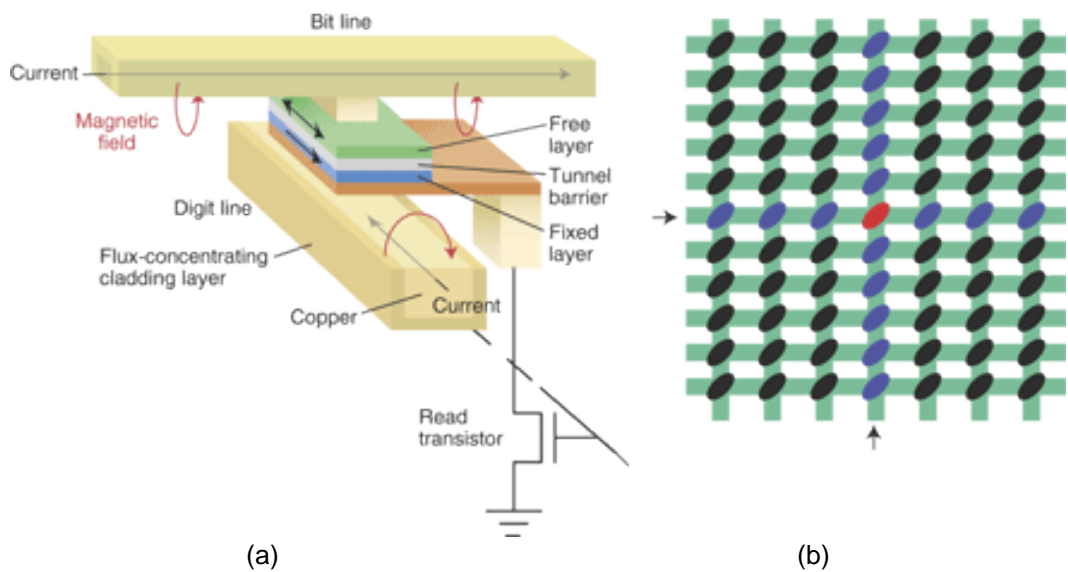

(a)                                                                  (b)

**Figure 18. (a) MRAM bit cell in series with a transistor for bit read selection. Perpendicular write lines above and below the MTJ select a single tunnel junction for writing; (b) Top view of an MRAM array, highlighting the fully selected bit (red) in the center and 1/2-selected bits (blue) along each current-carrying write line [AKE05].**

According to different switching mechanisms, MRAM can be classified into diverse categories [POR14]: (i) Field Induced Magnetic Switching (FIMS); (ii) Thermally Assisted Switching (TAS); (iii) Spin Transfer Torque (STT); and (iv) TAS-STT, which joint the two last switching mechanisms. STT-MRAM is considered as the most promising one due to its high-power efficiency and high switching speed [KAN14][POR14]. Here on after, this will be the technology chosen for MRAM discussion.

MRAM presents the following properties: (i) high write endurance when compared to other types of emerging non-volatile memory technologies; (ii) zero standby leakage power; (iii) fails to make simple higher levels of cache (L1 and L2) due to long latency and relatively high energy consumption; (iv) low chip yield and reliability, when compared to traditional CMOS technology; (v) low density, when compared to PCRAM; and (vi) higher write latency and energy consumption than SRAM [KAN14][KOM14][MIT14].

## 3.6  Domain Wall Memory (DWM)

DWM (or Racetrack memory) is a memory in which magnetic domains are used to store data in tall columns of magnetic material on the surface of a silicon wafer. This structure, illustrated in Figure 19, is used to build truly 3D devices and, hence, tackle additional approaches to the conventional means of developing cheaper and faster circuits, when compared to 2D transistors (such as DRAM) [PAR08].

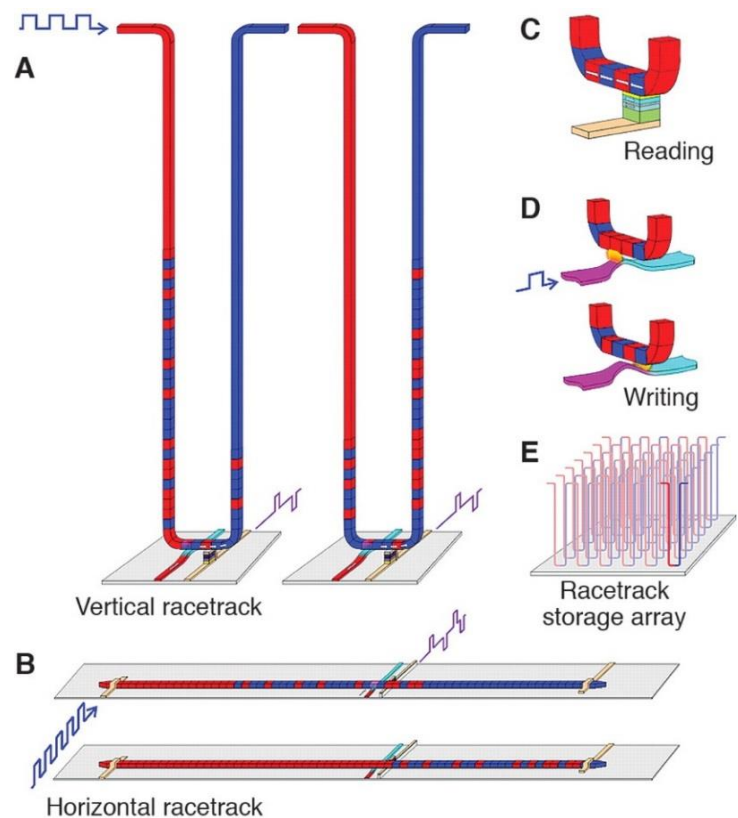

**Figure 19. (A) Vertical configuration racetrack; (B) Horizontal configuration racetrack; (C) Reading operation; (D) Writing operation; (E) Arrays of racetrack storage to form a high-density storage [PAR08].**

DWM integrates many magnetic pillars (called as magnetic domains) in one strip (called racetrack), which is connected to a few access transistors. Access to data information is obtained by shifting the magnetic domain along the racetrack and aligning the target domain to an access device. The writing process respects a similar shifting operation, meaning that DWM is a shift register. Accessing information depends on its location, which is relative to the access port and makes the performance dependent on the number of shift operations required per access. Thus, this type of memory presents several challenges to be used in the memory hierarchy [SUN14].

DWM presents the following properties: (i) Racetrack memory is highly dense, even when compared to other types of nonvolatile memories; (ii) outstanding write endurance; (iii) challenging design to compensate the required shifting operation in every read and write operation; (iv) challenging design to compensate high energy consumption of write operation [SUN14][VEN13].

## 3.7 Comparison of Memory Technologies

Table 2 presents a comparative evaluation of key differences on memory technologies. The items below aim to detail each row on this table.

**Maturity**: State of the art of the technology.

**Cell Size ($F^2$)**: Cell size is the necessary size for the basic cell unit of a given memory technology. Cell size is given in units of $F^2$, where F is the smallest lithographic dimension in a given technology. Normally, each cell contains one bit, however, racetrack memories like DWM can store multiple bits per cell [AMI13].

**Speed (R/W)**: Relative qualitative comparison of read and write operations, respectively.

**App. read time**: Approximate access latency of a read operation. The exact value of such operation is dependent on a given fabrication process.

**App. write time**: Approximate access latency of a write operation. As above, the exact value is dependent on a given fabrication process. PCRAM has an important disparate access latency of writing values (SET operation for value one and RESET operation for value zero) because writing in this memory must change the temperature of the chalcogenide [DON09b].

**Standby power**: Power consumed by a memory node that is idle; i.e., not performing read/write operation. Volatile memories must maintain their nodes active all the time, hence, this type of consumption can have significant impact in the overall system energy [LIU11].

**Read Energy**: Energy consumed in a read operation.

**Write Energy**: Energy consumed in a write operation.

**Write endurance**: Number of write/erase cycles that can be applied to a memory node before the storage media becomes unreliable, in average. PCRAM has the lower endurance of the analyzed memories, requiring strategic design to overcome this problem. For instance, Lee et al. [LEE09] propose partial writes (write only modified cache/word lines).

**Norm. density**: Normalized density to SRAM in 45nm technology.

**Data storage**: Physical element that store information in a given technology.

**Non-volatility**: The ability to retain data information when power is turned off.

**Table 2. Comparison of key differences of memory technologies[2].**

| | SRAM | DRAM | eDRAM | PCRAM | MRAM (STT) | DWM | References |
|---|---|---|---|---|---|---|---|
| Maturity | Product | Product | Product | Test chips | Test chips | Experimental | [MEE14][MUL04] |
| Cell size ($F^2$) | 120-200 | 6-8 | 60-100 | 4 – 12 | 6-50 | ≥ 2, 12 | [CHI15][FUK09] [MIT14][PRZ90] |
| App. read time (32nm) | 2-4ns | 2-6ns | 2-6ns | 2-6ns | 1-2ns | 1-3ns | [CHA13][DON09a] [LEE14][MAD09] [VEN12][ZHA14a] |
| App. write time (32nm) | 2-4ns | 2-6ns | 2-6ns | 40-46ns (reset) 100-110ns (set) | 2-5ns | 3-4ns | |
| Standby power | Leakage | Leakage & refresh | Leakage & refresh | Zero | Zero | Zero | [DON09a] |
| Read energy (32nm) | 0.10-0.80nJ | 0.60-0.80nJ | 0.50-0.70nJ | 0.10-0.70nJ | 0.06-0.20nJ | 0.08-0.60nJ | [CON13][DON09a] [VEN12][ZHA14a] [YU14] |
| Write energy (32nm) | 0.10-1.4nJ | 0.60-0.80nJ | 0.50-0.70nJ | 6-13nJ (reset) 2-6nJ (set) | 0.10-0.60nJ | 0.10-0.80nJ | |
| Write endurance | $10^{16}$ - $10^{18}$ | $10^{16}$ | $10^{16}$ | $10^8$ - $10^{12}$ | $10^{12}$ | $10^{16}$ | [DON09a][LEE14] [MIT14] |
| Norm. density (45nm) | 1 | 4 | 4 | 16 | 4 | 6 | [WU09][VEN12] [ZHA11] |
| Data storage | Latch | Capacitor | Capacitor | Chalcogenide | Magnetic | Magnetic | [CHA13] |
| Non-volatility | No | No | No | Yes | Yes | Yes | [DON09a][FUK09] |

[2] The values presented here are collected from related work, and may differ according to the manufacturing technology used in each foundry.

# 4  RELATED WORK ON 3D MPSOC ARCHITECTURES WITH ON-CHIP CACHE SUPPORT

This section presents a non-exhaustive list of current research effort on 3D MPSoC with cache hierarchy support bringing design possibilities, benefits and drawbacks.

Zhang et al. [ZHA14b] summarize current research effort in the design of 3D CMP, with tiers[3] dedicated for memory hierarchy. The survey focusses on two categories of architectures for 3D chip multiprocessors: stacking cache-only and stacking main memory.

Loi et al. [LOI10] present a 2D NoC with a vertically stacked memory layer. Each processor has fast access to a stack of memory banks on its top and remote slower access to memory stacks of other processors. The main contribution of this paper is the development of a 3D-DRAM controller responsible for administering the two types of memory accesses described above. Cadence SoC Encounter and Synopsys Design Compiler were used to obtain a synthesized platform. Experimental results show peaks of 4.53 GB/s for local memory access, and 850 MB/s for remote access through the NoC.

Fu et al. [FU14] propose a distributed memory with direct access to local and remote cache banks due to previous unsatisfactory results achieved with packet-based communication. In the distributed memory with direct cache access, the local cores access remote memory through remote-to-local virtualization without any network protocol translation. Every core has an auxiliary memory controller to access the local and remote memories. This controller is divided into two for managing the core communication and to handle the actual memory bank. Interconnection for memory accesses between cores is done using multiple buses. Two unidirectional buses are employed on each combination of column and row of a mesh topology. For an 8×8 topology, 16 buses are used. Each core in this architecture accesses a private 16KB L1 cache and a 64KB L2 cache. Simulation results with the PARSEC benchmark show that direct memory outperforms packet-based access in terms of both memory access latency and IPC by 17.8% and 16.6%, in average, respectively. The rationale for this is twofold: the lack of network protocol translation overhead in direct memory access and the reduced contention for using multiple independent connections. Gem5 simulator was employed to trace the memory accesses of both PARSEC and SPEC2006 benchmarks. Subsequently, they were fed to an in-house C++ cycle-accurate simulator responsible for simulating all implementation details of both direct and packet based communication.

Kim et al. [KIM13] describe a Massively Parallel Processor (MPP) with stacked memory called 3D-MAPS, which consists of a 64-core tier and a 64-memory block tier. Each core communicates with its dedicated 4 KB SRAM block. This chip was built with a two-tier 3D stacking technology using TSV and face-to-face bond pads. The estimated fabrication cost of 3D-MAPS compared to a theoretical 2D-MAPS (each memory block placed right besides its corresponding core) is approximately half the cost. The design and analysis of this chip was conducted using commercial tools from Cadence, Synopsys and Mentor

---

[3] In the 3D technology, a tier consists of a single 2D layer, and two or more tiers are stacked and connected to perform a 3D system.

Graphics, as well as in-house tools for handling 3D technology characteristics. They are currently working on a second version of this chip, called 3D-MAPS v2, which will be comprised of five tiers: two logic chips and three DRAM chips. In addition, they intend to double the core count for this version [GEO15].

Fick et al. [FIC13] present a large-scale 3D CMP with a cluster-based near-threshold computing architecture called Centip3De. A 3D stacking technology is used in conjunction with 130 nm CMOS. The 64 cores are organized into 4-core clusters and their aggregate cache is combined in a single shared 4x-larger cache. This larger cache has an increased voltage and frequency to assist all four cores. The use of an increased frequency when compared to the cores allows the cache to maintain single-cycle latency, even when the access is shared. Eight buses of 128 bytes assist the communication of all 16 clusters. The buses are split into two columns that span the cache and core layers of the chip. The system architecture was described and validated using the Gem5 simulator [BIN11]. Experimental results were conducted using the SPLASH-2 benchmark. The future version of this chip intends to employ a seven-layer chip: two of cores, two of caches and three of DRAMs.

Guthmuller et al. [GUT12] present a modular and scalable manycore architecture with multiple stacked cache tiles, as shown in Figure 20. In this architecture, each cache controller is responsible for a given segment of the main memory. At assembly time, multiple stacks of identical cache tiers can be incorporated into the architecture. In addition, at runtime, the cache quantity allocated to a memory segment can be tuned. The tier is comprised of cores with L1 and L2 caches interconnected by a local bus. The cache tier serves as a L3 cache. The SoCLib platform [SOC15a] was employed to perform experimental analysis of the proposed architecture. Two memory configurations were experimented with a SRAM cache and a mixed SRAM/eDRAM cache. Total area of the cache block is reduced by 47% with eDRAM/SRAM (compared to pure SRAM), since eDRAM presents higher degree of density. In addition, the authors demonstrated that the silicon area overhead due to die stacking (mainly, TSV area) can be as small as 10% of the total die area.

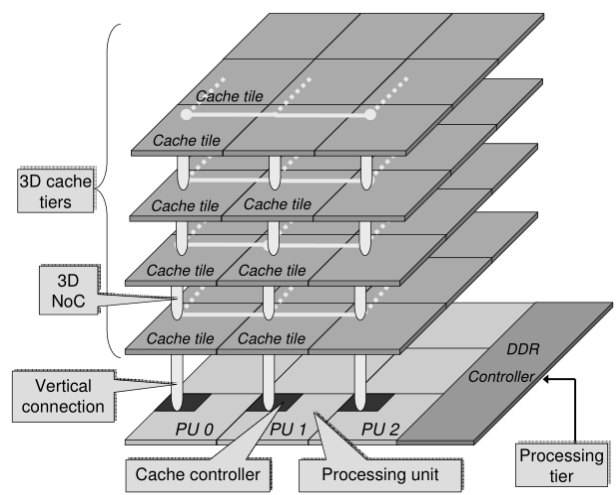

**Figure 20. Manycore architecture as proposed by Guthmutter et al. [GUT12].**

Li et al. [LI06] present a topology design mixing 2D NoC and TDMA bus. The 2D NoC interconnects nodes (i.e., CPU or caches) that are placed in the same layer. To eliminate the multi hop latency penalty inherently present in a packed based NoC, the TDMA bus interconnects the multiple stacks of layers. In addition, TDMA is used to eliminate the transactional nature of a distributed arbitration. Thus, a single-hop and transaction-less communication is achieved. The processors have private L1 caches and share a large L2 cache. Results conducted using the Simics simulator [MAG02] demonstrate that a 3D L2 memory architecture generates much better results than the conventional 2D design.

Wu et al. [WU09] describe two types of hybrid cache architectures: inter cache level, where the levels in a cache hierarchy can be made of disparate memory technology; and intra cache level, where a single level of cache can be divided into multiple segments, each one containing a different memory technology. The latter uses a fast segment for most accessed addresses and a slow segment for the remaining. This fast segment uses SRAM memory technology, because it presents the best latency of the four technologies analyzed. For the slow segment, three types of memory technologies are evaluated: eDRAM (volatile), MRAM (non-volatile) and PCRAM (non-volatile). The authors simulate a myriad of benchmarks, to show that an inter cache hybrid architecture design can provide 7% IPC improvement over a baseline 3-level SRAM cache design under the same area constraint. Those results were obtained using a PowerPC-based simulator called Mambo [BOH04]. Table 3 contains the summary of related work mentioned above.

**Table 3. Related work summary.**

| Work | Memory technology | L2/L3 present | Memory layer intended for | Requirements | Traffic | Full-system simulator |
|------|-------------------|---------------|---------------------------|--------------|---------|------------------------|
| [LOI10] | DRAM | No/No | Main memory | latency, area | JEDEC standard | - |
| [FU14] | DRAM | Yes/No | Main memory | latency, throughput | PARSEC/SPEC2006 benchmarks | In house |
| [KIM13] | SRAM | No/No | Cache memory | latency, throughput | Eight benchmarks | - |
| [FIC13] | SRAM | No/No | Cache memory | throughput, energy consumption | SPLASH-2 benchmark | Gem5 |
| [GUT12] | SRAM/eDRAM | Yes/Yes | Cache memory | scalability | SPLASH-2 benchmark | SoCLib |
| [LI06] | Not specified | Yes/No | Cache memory | latency, energy consumption, area | SPEC OMP benchmark | Simics |
| [WU09] | SRAM + (eDRAM/MRAM/PRAM) | Yes/Yes | Cache memory | latency, scalability | Various benchmarks | Mambo |

# 5   GEM5: FULL SYSTEM SIMULATOR

A full system simulator is a fast architecture simulator capable of executing software stacks from real systems (user and kernel code) without any modification [LEU10]. Such tool can create virtual platform designs that can gather experimental data with workloads compatible with the running software. A key feature of such simulator is the flexibility to explore architectural designs without the inherent hardware cost of doing so manually.

The simulation of computer architectures requires tremendous computational effort, since it is comprised of any number of processors, memories and I/O devices. To achieve an accurate hardware-level simulation, it is necessary low level descriptions, such as RTL (*Register Transfer Level*), and a detailed hardware simulation model, which increases the design exploration time making prohibitive the entire system simulation [BUT12][GUT14]. Therefore, simulators often use models of higher abstraction level that that exchange precision for efficiency. A few simulators allow the designer to choose the degree of precision desired. Hence, they can basically operate in two execution modes [BIN11][IBM07]: *simple (atomic)* and *cycle-based*.

*Simple (atomic) mode* enables to capture only the program execution, without regard to the timing accuracy. Resource contention is normally ignored, and a fixed latency is used instead. The operating system can be abstracted and, therefore, the simulator emulates each system call. Memory accesses are assumed synchronous and instantaneous. This model is intended for rapid software development/debugging due to its fast execution time.

*Cycle-based mode* enables to capture not only all the functionality of atomic mode, but also the accurate timing information. This mode supports resource contention through arbiters, queues and interrupts. This mode is intended for architecture exploration and platform design as it gathers information data with a greater level of fidelity.

This section describes the Gem5 simulator – its structure, flexibility and known limitations. Afterward, an overview of modern full system simulators and its distinctions is discussed.

## 5.1   Introduction to Gem5

Gem5 is a full system simulator that employs a flexible and highly modular discrete event model. This simulator is the result of the combined effort of a myriad of academic and industrial institutions such as AMD, ARM, University of Michigan, University of Texas and others. Currently, Gem5 supports six commercial ISAs (*Instruction Set Architectures*) (Alpha, ARM, MIPS, POWER, SPARC and x86) and boots the Linux Kernel on at least three of them (ARM, Alpha and x86) [BIN11]. The current state of those ISAs is maintained at the following location: [GEM15a]. Gem5 uses a BSD-like license that allows commercial and academic use and distribution of source code and binary formats [BIN11].

The goal of Gem5 is to be a community tool focused on object-oriented design for architecture modeling [BIN11]. Utilizing standard and message buffer interfaces, Gem5 follows a TLM (*Transactional Level Modeling*)-like semantic. This enable ample support for community-based changes on the simulator. [GEM15b] is the site where changes are proposed and reviewed by the community.

Gem5 supports *System-call Emulation* (SE) and *Full-System* (FS) modes. They mirror Simple (atomic) and Cycle-based modes described previously. The SE mode allows to handle the most commonly used system calls. Whenever the program requests a system call, Gem5 traps and emulates the expected result. In this mode, no effort is made to model devices and other operating system services. On the other hand, FS mode models a bare-metal environment suitable for running an OS. Because of the complexity of this mode, not all ISAs present in Gem5 can run it. Currently, Alpha, ARM, SPARC and x86 ISAs are supported [BIN11][GEM15a].

FS mode supports four different CPU models: *AtomicSimple, TimingSimple, In-Order* and *O3. AtomicSimple* and *TimingSimple* are non-pipelined CPU models that conduct the basic cycle of an instruction (fetch, decode and execute) and commits it on every cycle of execution. The *AtomicSimple* model is a single IPC (*Instruction Per Cycle*) CPU, which executes all memory accesses instantaneously. The *TimingSimple* model supplements the execution with the timing of memory accesses. Figure 21 shows the differences of execution cycle between these two models. *In-Order* and *O3* (*Out-of-Order*) are "execute-in-execute" CPU models that emphasize instruction timing and simulation accuracy. "Execute-in-execute" means that instructions are executed only in the execute stage of the pipeline. While *In-Order* is restricted to execute instructions in the order that they are received, *O3* models an out-of-order CPU. Both models have parameterizable resources like the number of pipeline stages, load/store queue and reorder buffer (*O3* model only). The challenge of using these two last models is the simulation time required. They are roughly an order-of-magnitude slower than the simpler models [SAI12]. In addition, not all ISAs support those detailed models. The ARM ISA, for instance, has *AtomicSimple, TimingSimple* and *O3* models but lacks the *In-Order* model [END14].

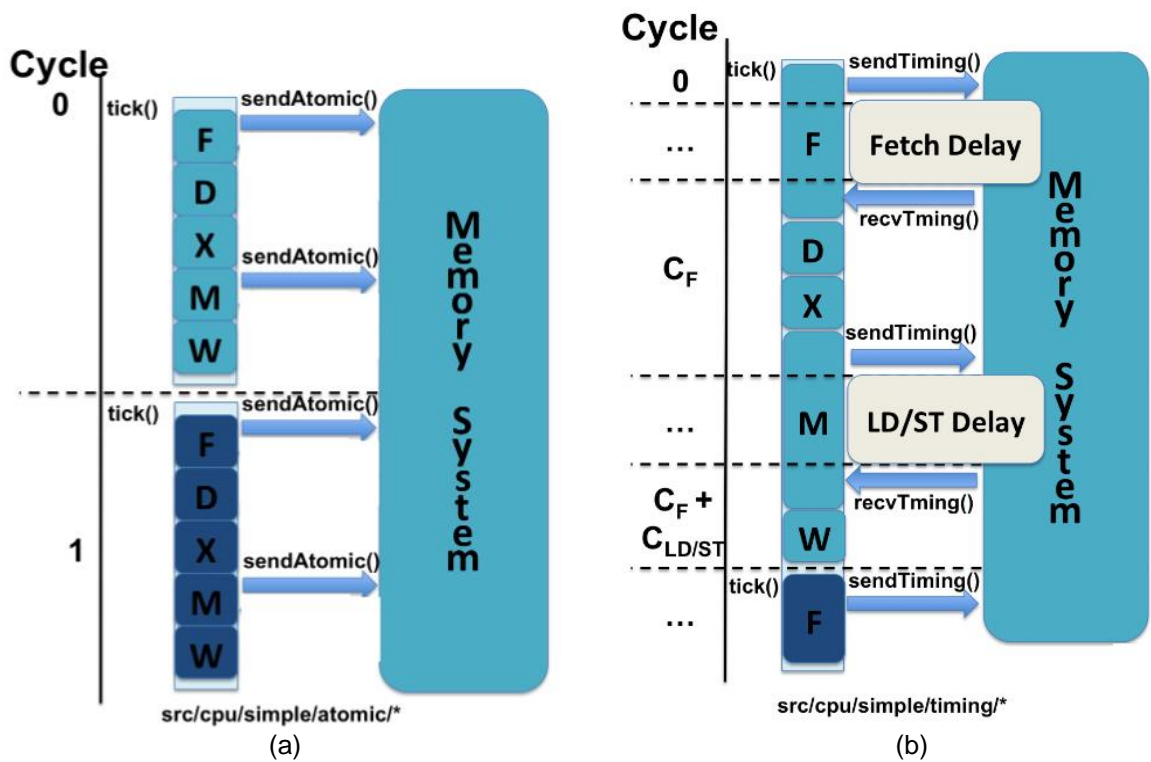

**Figure 21. Schematic of (a) AtomicSimple and (b) TimingSimple CPU models [SAI12].**

Two different memory systems are present in Gem5: *Classic* and *Ruby*. The *Classic* mode was inherited from the m5 simulator [BIN06], while *Ruby* was inherited from the gems framework [MAR05]. Classic mode is the faster of the two and provides ease configuration through python scripts. Currently, this mode maintains memory coherence through a MOESI-like snooping protocol [SUH06]. Changing this protocol requires an overhaul of the entire cache system. In contrast, *Ruby* sacrifices simulation speed to provide a flexible infrastructure to simulate a wide variety of memory systems. In particular, *Ruby* provides a specific language, where one can define more easily cache coherence protocols. In addition, this mode supports many interconnection networks such as crossbar, mesh and point-to-point topologies. Unfortunately, *Ruby* is limited to Alpha and x86 ISAs [GEM15c].

Devices in Gem5 are built on a base class called *io_device*. The device must define three fundamental functions from this class: getAddressRange, read and write. The getAddressRange function return the address range the device responds to. This information must be provided for the core simulation engine. The read and write operations are performed in their respective functions, so that the device can interact with the remainder of the system. A number of devices are already implemented in the Gem5 framework. Example of such devices are Network Interface Controllers, Hard disk controller, DMA (*Direct Memory Access*) engines, UART (*Universal Asynchronous Receiver/Transmitter*) and others [BIN11][GEM15d].

## 5.2  The Accuracy of the Gem5 Simulator

Accuracy is one of the key aspects presented in Gem5. This attribute is intended to be balanced in regard to the simulation speed desired by the user [BIN11]. Therefore, the user has some control of the accuracy presented in this simulator. Since much of the source code is developed by academia studies, the focus of development is many times to study new concepts instead of replicating existing hardware modules. For instance, Wiener [WIE12] incorporates the ARM CoreLink CCI (*Cache Coherent Interconnect*) into the Gem5 simulator. However, this is not a simple port – while CCI is aimed at single-hop interconnections, the proposed Gem5 counterpart supports multi-hop interconnections. On the other hand, simplifications were made to the underlying coherence protocol. One example of this is the snoop hit scenario. When a snoop hit occurs, the memory controller does not handle the request, instead, it destroys the snoop hit immediately. This means a speculative fetch to the primary memory is "magically" avoided [WIE12]. Nonetheless, Gem5 still aims to accurately model state-of-the-art systems. Recently, studies were conducted to evaluate this crucial goal of Gem5.

Butko et al. [BUT12] were one of the first published paper that discusses Gem5 accuracy in terms of performance estimation. They used the Snowball SKY-S9500-ULP-C01 development kit as the reference hardware model. This development kit comprises a dual-core ARM Cortex-A9 processor. Experimental results showed that the mismatch between the real hardware and the simulation system range from 1.39% to 17.94%. The benchmarks employed were selected applications of the SPLASH-2 and APLBench suite. The primary reason for the discrepancy encountered in their work is the abstraction used in the model of the external DDR memory latency.

Endo et al. [END14] propose an In-Order CPU model for the ARM ISA based on the O3 model of the same ISA. With both models, the timing accuracy of Gem5 is evaluated with real hardware comparing the execution time of 10 benchmarks of PARSEC 3.0. The Cortex-A9 model (O3 CPU model) estimates the execution time with an absolute error of only 7.4% (ranging from 1% to 17%), in average. The In-Order model (Cortex-A8) estimates the execution time with an absolute error of 8% (ranging from 2% to 16%), in average. The authors conclude that, even considering the generic nature of Gem5, the magnitude of error encountered can be considered good for an architecture simulator.

Gutierrez et al. [GUT14] investigate source of discrepancies in latency estimation between the Gem5's ARM ISA and the execution of a real hardware platform (ARM Versatile Express TC2 development board). Only the O3 CPU model was tested in this work. Using the PARSEC benchmark, it was observed an average of 11% and -12% of runtime deviation, for single and dual-core systems, respectively. The work also shows that when measuring multi-threaded benchmarks, Gem5's scaling is accurate to within 1%, in average. Changes on the Gem5 were proposed and submitted to the Gem5 source code to reduce the overall inaccuracies encountered.

All the works cited here agree that the discrepancies are within acceptable range, since Gem5 supports many ISAs and does not have a commercial nature. Nonetheless, it is important to understand these deficiencies and mend them whenever it is possible.

## 5.3  Simulation Design and Flow

The core of Gem5 simulator is an event-driven engine, which tightly combines C++ and Python programming languages. Every component in the simulation is represented simultaneously as a C++ object and as a Python object [WIE12] to enable effortlessly composition of any system. The designer needs to recompile the platform only if changes the behavior of some components or if he wants to increase/decrease the level of verbosity/optimization of the simulator. Otherwise, the platform can be done just by editing Python scripts.

Figure 22 shows the initialization process of a "Hello, World" example running in SE mode. The first code (C++) is the main function, where the user can define some variables (e.g., debug flags), invoke the python debugger, enable remote gdb debugging and so on. Following, the designer invokes a python script that describes all objects to be instantiated and the way they are interconnected – in other words, the architecture is defined here. The designer can also define variables to this script, changing the architecture. An example of Gem5 invocation of an ARM platform in FS mode is: *./build/ARM/gem5.opt –debug-flags=Ethernet,IdeDisk configs/example/fs.py –mem-size=512 –caches –num-cpus=4*.

Interaction with the system is twofold: telnet connection and VNC (*Virtual Network* Computing) session. For the telnet connection, Gem5 provides a specialized program called m5term. For VNC session, the user must use an external application. Telnet is limited to keyboard interactions only, whereas VNC enables additional mouse inputs [WIE12].

Gem5 also supports checkpoints in the system, so that a user can start execution at his desired region of interest. This enables the user to fast-forward large workloads that can take many hours just to initialize. This mitigate some of the slower execution of Gem5 ISS (*Instruction Set Simulator*) when compared to binary translation-based simulators [BIN11].

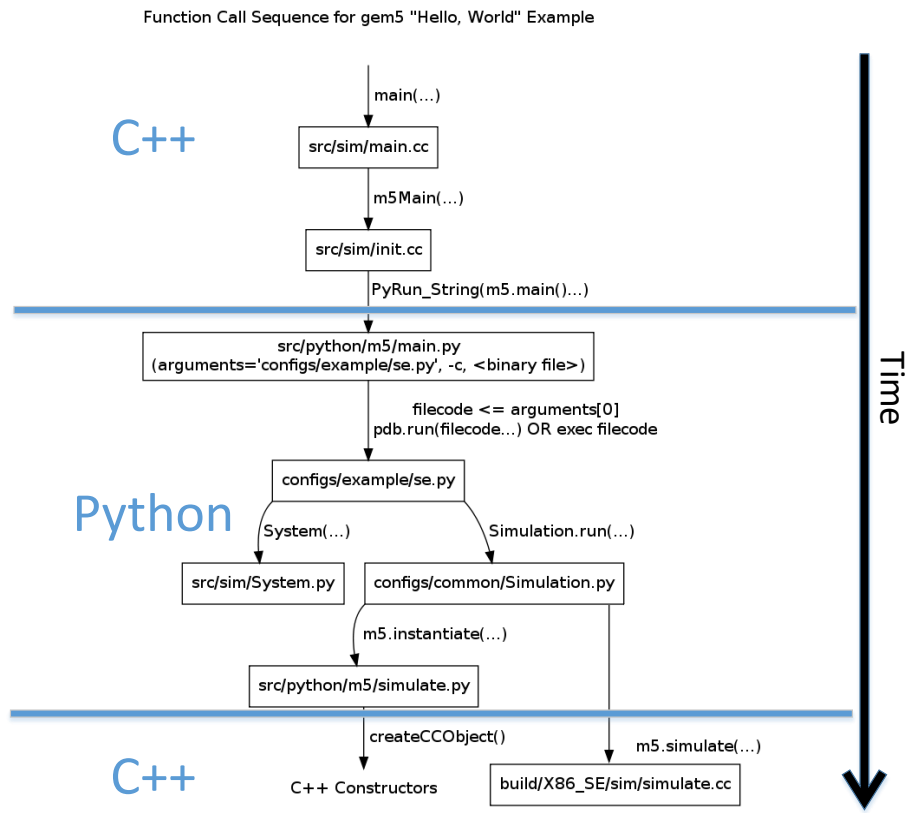

**Figure 22. Initialization of Gem5 [GEM15e].**

Another interesting feature to mitigate the slower execution of detailed CPU models is a program called M5ops. This program enables special instructions on the executing simulation to trigger simulator events. Two functions are especially useful: dumpstats and switchcpu. Dumpstats clears all the simulation statistics until its call. This is useful for cleaning the warming up process of the system. Switchcpu causes the simulation to quit with an event of type "switch cpu". The user, then, can check for this type of event and change the CPU model for a different kind. In this way, the system can execute faster until the region of interest and only then changes its model for a slower one. Gebhart et al. [GEB09] demonstrate the compiling and executing process of the PARSEC benchmark using such features on Gem5.

Gem5 generates a couple of files after the end of the simulation. Figure 23 depicts simulation input, runtime interfaces and output. Simulation is ended either by the user or by choosing a maxtick parameter. In FS mode, the following outputs are produced:

**Simout and simerr** - The standard output and error stream generated by the simulated OS.

**System.terminal** - The output of the simulated system's terminal.

**Framebuffer.bmp** - The latest contents of the simulated system's display.

**Config.ini** - A key output of the Gem5 simulator. This file shows all components instantiated, its interconnections and its respective parameters. This allows the validation of the system simulated with the one intended by the user.

**Stats.txt** - The second key output of the Gem5 simulator. This file aggregates all statistics generated by every component in the system. The overall extent of this file is limited by the implementation of the system's components. Fortunately, the components already implemented in Gem5 have a good amount of statistics gathering.

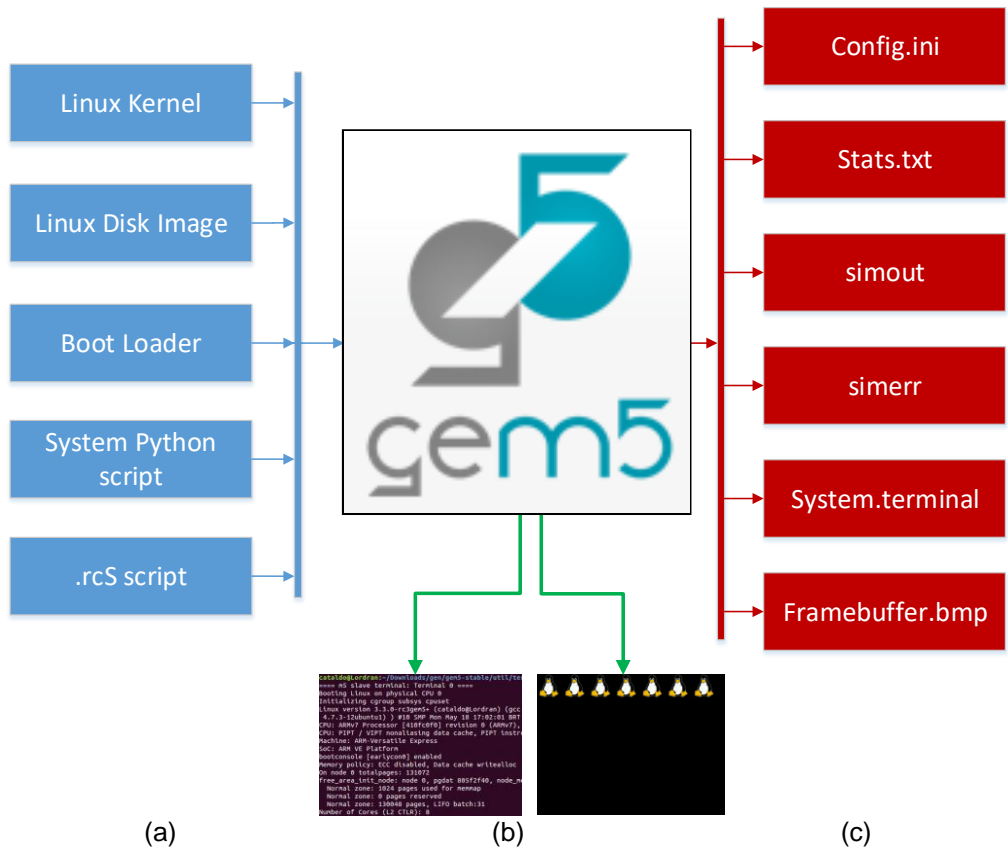

(a)                                        (b)                                        (c)

**Figure 23. Gem5 simulation (a) inputs, (b) runtime interfaces and (c) outputs (Based on [WIE12]).**

## 5.4  Overview of Full System Simulators

This section presents an overview of some relevant full system simulators that are employed in the design space exploration of MPSoC platforms. The criterion for choosing these simulators was established in accordance of published works in areas related to design exploration of MPSoC and/or cache.

### 5.4.1  SoCLib

SoCLib [SOC15a] is an open platform for virtual prototyping of MPSoCs described in SystemC language, providing high-level abstraction while maintaining accurate transaction-level results. In this platform, processors are described using ISS. Currently, SoCLib is maintained at Lip6 laboratory in France. SoCLib is licensed under GNU's Not Unix (*GNU*) General Purpose License (*GPL*) version 2.

Hardware is implemented using one of two types: TLM-DT (*TLM with Distributed Time*) or CABA (*Cycle Accurate Bit Accurate*). TLM-DT is a model compliant to TLM2.0 OSCI (*Open SystemC Initiative*) standard [SOC15b]. However, they differ on the time representation, because an absolute time is used instead of the global simulation time provided by the SystemC core. In addition, all messages are annotated with timing information, since synchronization between timed processes are no longer centralized [TEC15]. CABA aims to model hardware at the cycle accurate level. As stated in [SOC15c]: "*The idea is to force the 'event driven' SystemC simulation engine to run as a cycle-based simulator*". To do this, a hardware description is taken and converted to three types of function: (i) **Transition function**, which is responsible for the computation of the next value of registers. It takes as input the current values of the registers and input signals; (ii) **Moore generation function**, which is responsible for the computation of output signals that only depend on the internal registers; and (iii) **Mealy generation function**, which is responsible for the computation of output signals that depend on the internal registers and on the values of the input signals.

Figure 24 shows an example of a simple hardware and its representation using a graph, which is the input of the CABA conversion algorithm that produces the description of Figure 25. Fraboulet et al. [FRA04] defines this and an extended algorithm to avoid unnecessary duplicated code. As expected, this higher-level of detail compared to TLM-DT results in an expressive increase of simulation time – CABA is approximately 10 times slower than TLM-DT [POU09].

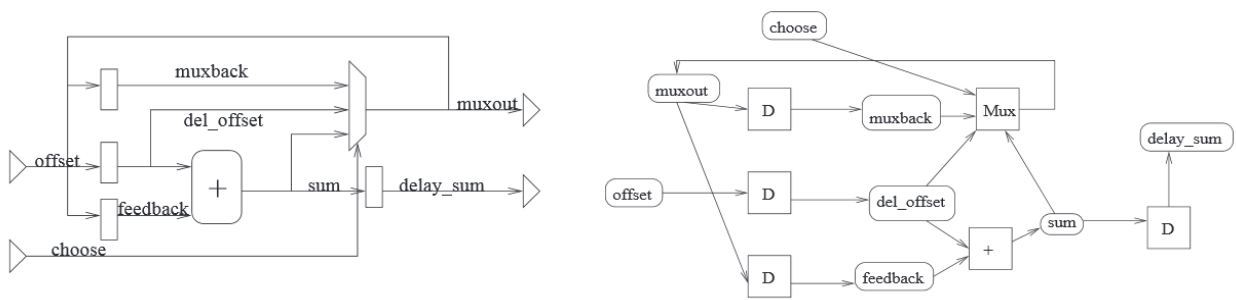

**Figure 24. Example of a simple hardware and its representation as a graph [FRA04].**

The most recent SoCLib version supports four OSs [SOC15a]. We chose to test MutekH OS, since it is maintained by the same laboratory, and we found many errors in SoCLib during simulation. Observing the repository activity, we can hypothesis that although SoCLib continues to be updated, the platform used for MutekH did not receive the same maintenance. There is a two-year gap and approximately 300 commits between updates in those two cases[4].

---

[4] Revision 2624 (SoCLib) and Revision 2325 (MutekH platform examples).

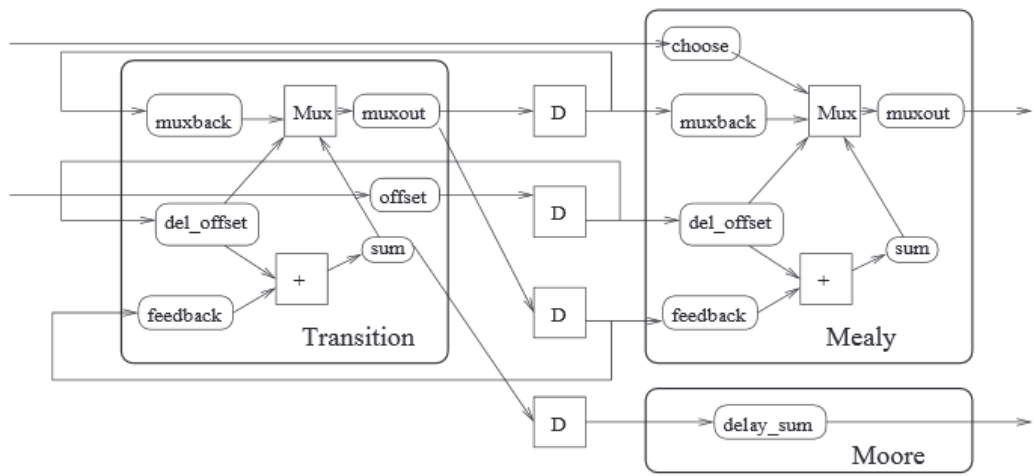

**Figure 25. CABA representation of the previous hardware description in Figure 24 (this is not the final optimized version) [FRA04].**

### 5.4.2 RABBITS

Rabbits is a system simulator that relies on QEMU for software execution and SystemC for hardware modeling [RAB15]. Architecture support is limited to the ARM family. QEMU is employed for its binary translation technique, as it improves the simulation time required for hardware/software evaluation [GLI09]. Figure 26 shows an example of platform in this simulator, whereas the processing units are executed in the QEMU framework and its communication with the outside system is wrapped in SystemC. Rabbits is licensed under GNU GPL version 3.

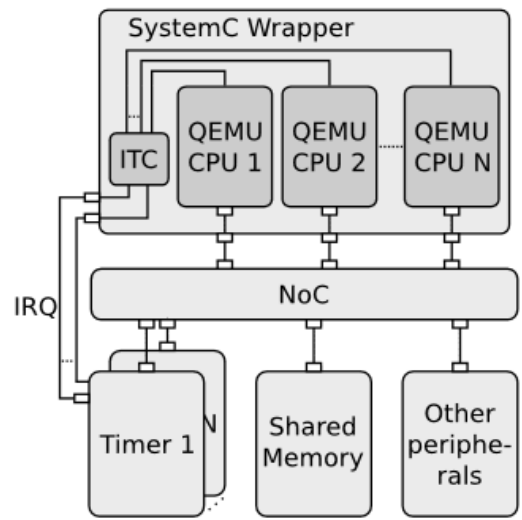

**Figure 26. Example of a simulation platform in Rabbits [GLI09].**

QEMU lacks any implementation of cache models [GLI09][MAH13]. Recently, there have been extensions for cache support in QEMU – Mahmoodi et al. [MAH13] proposed an extension for the MIPS architecture and Dung et al. [DUN14] for the ARM architecture.

However, Rabbits was designed earlier than this and rolls out its own implementation of cache in SystemC.

The major drawback of Rabbits is its lack of proper documentation. This is even alluded at its main page [RAB15]. Rabbits offers seven platforms in its repository. However, without documentation it is rather difficult to determine their characteristics. Because of this limitation, Rabbits was rejected for this work.

### 5.4.3 SIMICS

Simics is a full system simulator intended to create executable specification for hardware designers and a fast-virtual platform for software developers [WIN10]. Simics uses ISS for processor execution [MAG02]. Contrasting to the previous simulators described in this section, Simics is a commercial product and has a closed license [GUT14]. Access to its source code is handled on a case-by-case basis [WIN15a].

Hardware designers write TLM-like code using a proprietary language called Simics DML. This language uses a C-like programming syntax. The hardware is not analyzed to determine its delay; instead, the designer issues latency at a later stage [WIN10]. Figure 27 shows an example of a hardware interface with interrupt support.

```
bank regs {
    parameter register_size = 4;
    register version        @ 0x00 "Device version register";
    register control        @ 0x04 "Device control register";
    register status         @ 0x08 "Device status register";
    register reset          @ 0x0c is (write_only) "Reset register (write only)";
    register irq_num        @ 0x10 is (read_only) "IRQ assigned to device";
    register rule_set       @ 0x14 "Rule set (bit encoded)";
    register line_length    @ 0x18 "Line length (in bits)";
    register start_compute  @ 0x1c "Start computation";
    register input[32]      @ 0x20 + 4*$i is (write_only) "Input buffer";
    register output[32]     @ 0xa0 + 4*$i is (read_only)  "Output buffer";
}
```

**Figure 27. Example of a DML device programming interface [WIN10].**

For software developers, Simics offers a customized version of the widely used Eclipse IDE (*Integrated Development Environment*). In addition, virtual platforms are executed using the following operating systems: vanilla Linux, Wind River Linux, VxWorks and others [WIN15b].

Due to the lack of cache model and a memory latency model [WIN10], we decided to work on a different full system simulator.

### 5.4.4 OVP

Open Virtual Platforms (OVP) is a system simulator that uses dynamic binary translation to cope with system design complexity while still maintaining high simulation speed [REK13]. OVP has a dual license model [IMP13]: its main simulation core and some processor models are proprietary, while other processor models and APIs (*Application Programming Interface*) are open source via a modified Apache 2.0 license. OVP is comprised of three crucial components:

- **OVPsim**, which is the simulation engine, responsible for translating the target architecture binary code to x86 host instructions. It can be wrapped and called from other simulators environments like, for instance, SystemC [IMP15a]. While OVPsim is freely available (for non-commercial usage), CpuManager is the commercial alternative provided by Imperas. Some functionalities are exclusive to the commercial tool [IMP14].

- **Library of processor models** contains open-source and pre-compiled models of processing units. These models support various I/O components (e.g., UART and DMA) and memory systems. In regard to operating systems, Linux, Android and µcLinux are supported [IMP15b].

- **OVP APIs** are four interfaces for the C language. These interfaces are responsible for instantiation of full systems, creating new processing unit models and creating new peripheral models [IMP15c].

OVP is a software virtual platform that does not model hardware in a latency-aware manner [IMP11]. Consequently, it is designated as an instruction accurate simulator. Therefore, for a given processing unit, OVP guarantees that registers hold the correct values at the end of each instruction. However, there is no concern for pipeline progression, out of order execution or delays in the memory system [AGR09].

Some of the processor models support L1 cache [IMP15d]. However, the designer cannot customize the cache behavior. For this, the designer must roll out his own cache model implementation at the cost of simulation performance, as OVP is unable to optimize this scenario [IMP15d]. One of the creators of OVP, James Kenney, states that: "(…) *in terms of performance, on my 3Ghz PC, I expect to see several hundred MIPS simulation speeds for simulations without caches* (…) *and 10-20 MIPS when I have full MMCs, although this is of course highly dependent on the complexity of the MMC model* (…)" [IMP15e].

A challenge of using OVP for cache evaluation is its lack of proper latency model, since it assumes a 'perfect' memory model – where there is no latency penalty [IMP15f][IMP15g]. Therefore, OVP was rejected for this work since it is intended for software virtual platform, whereas this work aims to explore software/hardware virtual platforms. Nevertheless, restructuring OVP is undesirable since there are alternative simulators better suited for this scenario.

### 5.4.5 TAXONOMY OF FULL SYSTEM SIMULATORS

Table 4 depicts the key characteristics of the full system simulators discussed in this work. The items below describe the characteristics analyzed here.

**ISA(s) supported** in the simulator does not differentiate if the ISA is limited to atomic mode only or supports both modes of execution (atomic and cycle-based).

**Processor emulation** technique employed by the simulator. Binary translation is faster than ISS; however, it requires a complex implementation code [BIN11].

Primary **License** of the simulator. Note that some parts of the simulator (especially external programs) may employ a disparate license.

**Accuracy** employed by the simulator. Functional accuracy is limited to the behavior of the system, while cycle-accurate aims to detail the timing behavior as well.

**Operating systems** that are supported by the simulator. This may be restricted to a limited number of the overall ISA(s) supported by the simulator. Again, this is not differentiated in this table.

The presence of a **cache model**. This presumes the following operations: storage, coherence protocol and timing behavior.

**Table 4. Taxonomy of full system simulators.**

| Simulator | ISA(s) supported | Processor emulation | License | Accuracy | Operating systems | Cache model |
|-----------|------------------|---------------------|---------|----------|-------------------|-------------|
| SoCLib | SPARC, Nios II, POWER, MIPS, ARM and others | ISS | GNU GPL v2 | Cycle-accurate | DNA/OS, MutekH, NetBSD and others | Yes |
| Rabbits | ARM | Binary translation | GNU GPL v3 | Cycle-accurate | Linux and DNA/OS | No |
| Simics | x86, ARM, M68k, MIPS, POWER, SPARC, Alpha | ISS | Closed | Functional | Linux, NetBSD, Solaris, Windows, and others | No |
| OVP | ARM, MIPS, x86 | Binary translation | Dual license | Functional | Android, Linux, and others | No |
| Gem5 | POWER, ARM, MIPS, Alpha, SPARC, x86 | ISS | BSD | Cycle-accurate | Android, FreeBSD, Linux, Solaris | Yes |

# 6 DESIGN AND EXPLORATION OF 3D MPSOC ARCHITECTURE

This work explores 3D MPSoC architectures with on-chip cache support targeting disparate solutions. It defines multiple architectures with diverse tradeoffs, in order to evaluate the following constraints in the memory hierarchy: energy consumption, heat dissipation, latency and write endurance. Experimental analysis will be conducted using the Gem5 full system simulator.

Energy saving has become one of the most important design challenges as technology has advanced [RET11][TRA10]. Several previous works [BEN13][GOR07] have shown that more than 50% of energy consumption in processor-based architectures is consumed by the cache subsystem. Aiming to reduce energy consumption while maintaining acceptable performance requires careful design. Two categories will be explored in this domain: *Dynamic* and *Standby* energy consumption. In the context of cache subsystem, Dynamic energy consumption is the energy required to read ($Energy_{read}$) or to write ($Energy_{write}$) an information. Additionally, the entire dynamic energy consumption has to take into account the quantity of reads ($n_{read}$) and writes ($n_{write}$). Standby energy consumption is the power dissipated ($Power_{idle}$) during all time the system is idle ($time_{idle}$). Consequently, the total energy consumed by a given cache level may be modeled by Equation 1.

$$Energy = n_{read} \times Energy_{read} + n_{write} \times Energy_{write} + time_{idle} \times Power_{idle} \qquad (1)$$

Heat dissipation is a challenge in MPSoC, since it presents multiple systems executing into a single IC. Intense and continuous heat applied to an IC can damage devices and produce hardware faults [AMB11]. 3D stacking exacerbates this problem by increasing heat dissipation and thermal resistance [BRU09]. However, as shown in Figure 28, not all systems in an IC produce the same amount of heat and, hence, this can be exploited to design 3D heat aware architectures.

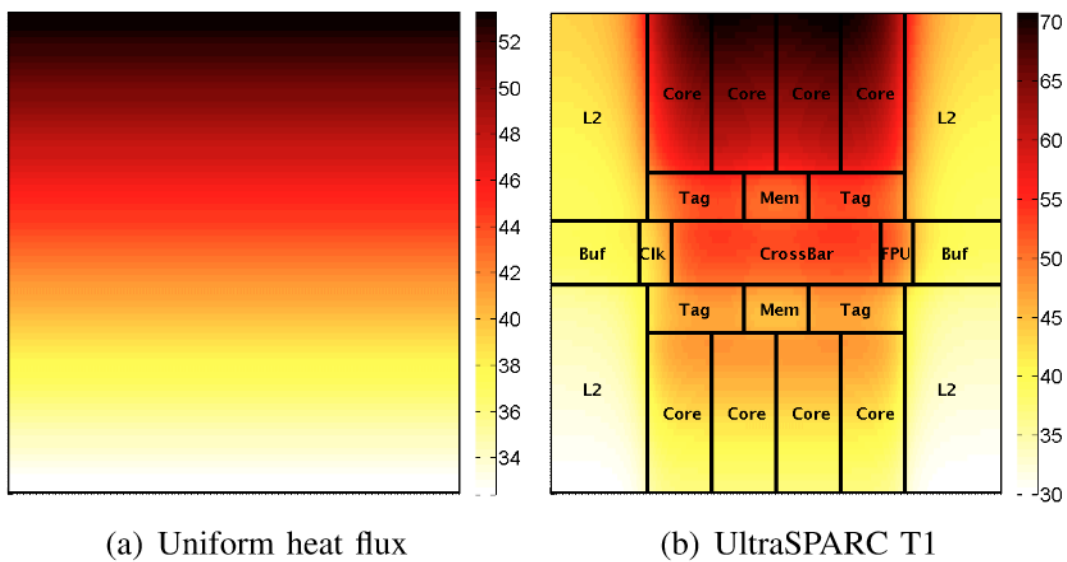

(a) Uniform heat flux          (b) UltraSPARC T1

**Figure 28. Temperature distribution of a two-die 3D IC with (a) uniform heat flux density of 50W/cm² and (b) the UltraSPARC T1 (Niagara-1) architecture [SAB12].**

Latency minimization is a common requirement in applications targeting on-chip architectures [BJE06][SAB10]. Unfortunately, such minimization can induce increase in other requirements such as area and energy consumption minimization [BEN13]. Therefore, a multi-constraint-oriented approach is recommended.

Efficient approach for memory written is an important requirement when considering the use of emerging memory technologies, because some of them do not have the same high endurance as traditional memory technologies. As such, some emerging memory technologies are inappropriate to be applied on some "write-bounded" application. Table 2 shows the endurance of many memory technologies.

The requirements discussed here are outlined in Table 5 for three levels of cache. The importance applied to each row is relative to the cache subsystem only.

**Table 5. Comparison of importance of some memory requirements for levels of cache. The colors indicate the importance of each level against the specified criteria. Green, yellow and red means low, moderate and severe importance, respectively.**

|   | Dynamic energy saving | Standby power saving | Dissipation of heat (in use) | Dissipation of heat (standby) | Latency reduction | High endurance |
|---|---|---|---|---|---|---|
| L1 | Severe | Moderate | Severe | Moderate | Severe | Severe |
| L2 | Moderate | Moderate | Moderate | Low | Moderate | Moderate |
| L3 | Low | Severe | Low | Low | Low | Moderate |

The basic principles of locality (temporal and spatial) that justify the use of caches [PAT13] are applied to the importance of each requirement. The principle of locality is described in the items following.

**Temporal locality**: If a particular memory location is referenced, then it is likely that the same location will be referenced again in the near future memory access.

**Spatial locality**: If a particular memory location is referenced, also adjacent memory locations tend to be referenced.

Those principles can be easily associated with two basic elements of programming languages: loop statements and sequential execution [LUT13]. Therefore, due to the principle of locality as new levels of caches are added, the following statements are made:

1. The importance of dynamic energy saving, and standby power dissipation minimization are inversely and directly proportional to the inclusion of new levels of cache, respectively.

2. The importance of heat dissipation is inversely proportional to the inclusion of new levels of cache.

3. Latency reduction and high endurance requirements are inversely proportional to the inclusion of new levels of cache.

## 6.1 Architecture Baseline

The architecture exploration starting point is the ARM system modeled by the Gem5 simulator. The ARM ISA was chosen for three reasons. Firstly, it is one of the ISAs supported by the Gem5 simulator. Secondly, it is widely employed in embedded systems such as mobile cellphone, automobile vehicles and developments kits [ARM15e][NVI15][SAM15b]. Thirdly, ARM is committed in helping the development of the Gem5 simulator and provides tool support for analyzing the ARM system behavior [ARM15f].

The ARM system of Gem5 is based on the ARM Versatile Express development board [GUT14]. Table 6 details the most relevant characteristics of the processor, cache and memory subsystems, respectively; and Figure 29 depicts a development board based on this architecture.

**Table 6. Characterization of the Versatile Express development board (Based on [ARM11a][ARM12a][ARM12b][ARM15a]).**

| Processor subsystem | Big.LITTLE architecture | |
|---|---|---|
| | 2x Cortex-A1 | 3x Cortex-A7 |
| Core Type | Out-of-Order | In-Order |
| Speed | 1Ghz | 800Mhz |
| Pipeline | 15 stages (integer) | 8 stages (integer) |
| Extensions | VFP & NEON | VFP & NEON |
| **Cache Subsystem** | Cortex-A15 | Cortex-A7 |
| L1 cache | Private L1 cache (32Kb instruction and 32Kb data) | |
| L1 I/D associativity | 2-way | |
| L1 I-cache block size | 64 Bytes | 32 Bytes |
| L1 D-cache block size | 64 Bytes | |
| L1 I/D replacement policy | LRU (*Least Recently Used*) | Pseudo random |
| L1 I-cache addressing | PIPT | VIPT |
| L1 D-cache addressing | PIPT | |
| L1 coherence protocol | MESI | MOESI |
| L2 cache | Shared L2 cache (1Mb) | Shared L2 Cache (512Kb) |
| L2 associativity | 16-way | 8-way |
| L2 block size | 64 Bytes | |
| L2 replacement policy | Pseudo random | |
| L2 addressing | PIPT | |
| L2 coherence protocol | MOESI | |
| Interconnection | Internal CoreLink CCI-400 | |
| **Memory Subsystem** | | |
| DRAM type | DDR2 x32 | |
| DRAM frequency | 400MHz | |
| Memory size | 2GB | |
| Memory Interfaces | 1 | |
| System bus frequency | 500Mhz | |
| **Components** | | |
| NOR Flash, UART, SD Card Controller, 10/100 Ethernet, HDLCD and others | | |

This board uses the big.LITTLE architecture, which means that it has a high-performance cluster of out-of-order processing units and a low-power cluster of in-order processing units. Both have the same extensions: ARM Vector Floating-Point (VFP) and a general-purpose engine called NEON for accelerating multimedia and signal processing algorithms [ARM15b]. On the cache subsystem, 2-way set associative is used for L1 caches and 16-way and 8-way are used for L2 caches of the high-performance and low-power clusters, respectively. Only the L1 instruction cache of the Cortex-A7 core uses VIPT (*Virtually Index Physically Tagged*), while all the other caches use PIPT (*Physically Index Physically Tagged*). For the high-performance cluster, the L1 cache employs the MESI coherence protocol, while the L2 cache employs the MOESI protocol. A dedicated hardware called SCU (*Snoop Control Unit*) coordinates the translation between the two protocols [ARM11a]. For the low-performance cluster, all caches employ the MOESI protocol. Finally, the memory subsystem is comprised of a 2GB 32-bit DDR2 clocked at 400 MHz.

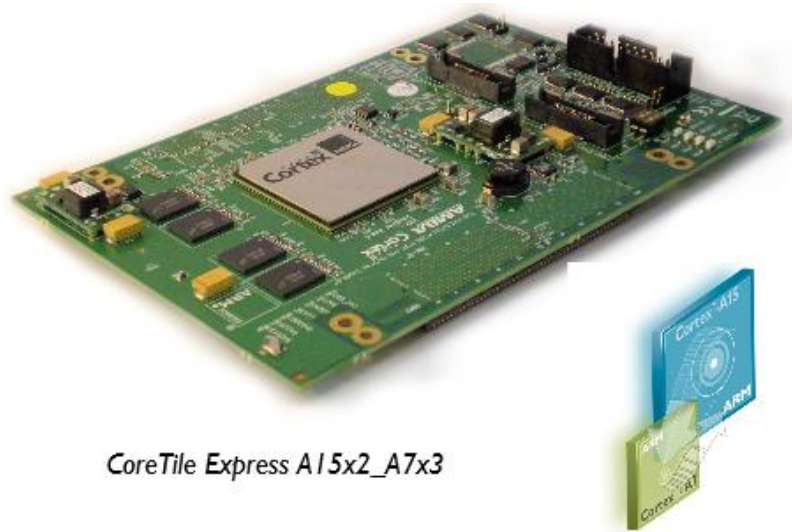

**Figure 29. Versatile Express development board [ARM15a].**

The Gem5's representation of this architecture is shown in Table 7. Internally, this architecture is called *VExpress_EMM*. Note that this table is limited to the default parameters and to the specific version of Gem5[5]. Basically, all parameters shown in Table 7 can be changed using Python scripts – the exceptions are: adding/removing pipeline stages, addressing policy employed in caches, coherence protocols employed in caches and the CCI. For these, Gem5's capabilities must be expanded.

In comparison to the Versatile Express board, there are discrepancies in all subsystems analyzed. In the processor subsystem, the speeds of cores are lower. However, this can easily be changed. In the cache subsystem, all caches use PIPT for addressing, as does Cortex-A15 (Cortex-A9 uses VIPT and PIPT for L1 and L2 cache addressing, respectively [ARM10]). Again, all caches use MOESI for cache coherence. Our hypothesis is that it happens because there is no SCU implementation on Gem5. Therefore, there is no hardware to coordinate between two coherence protocols. Interconnection is done through an expanded CoreLink CCI structure that supports multi hops networks. Wiener [WIE12] was the author responsible for this implementation and details its features and its shortcomings. For the memory subsystem, the default option is the DDR3 x64. Nevertheless, there is a model for DDR2 x32 that resembles the Versatile Express main memory. Finally, some components of the system are missing (SD Card Controller) and others have a broader range of operation (Hard Disk Controller).

---

[5] In this work, Gem5's version is identified by the following id: a48faafdb3bf (Mercurial repository identification). Revision 10666 from the Gem5 stable repository. Date of commit: February/2015.

**Table 7. Default parameters for the Gem5's ARM Versatile Express model (Based on [END14][GEM15f][GUT14][SAI12]).**

| Processor subsystem | Cortex-A9 based (ARMv7-A profile) |
|---|---|
| Core Type | Out-of-Order |
| Speed | 500Mhz |
| Pipeline | 7 stages (integer) |
| Extensions | VFP & NEON |
| **Cache Subsystem**[6] | Cortex-A9 based |
| L1 cache | Private L1 cache (32Kb instruction and 32Kb data) |
| L1 I/D associativity | 2-way |
| L1 I-cache block size | 64 Bytes |
| L1 D-cache block size | 64 Bytes |
| L1 I/D replacement policy | LRU (*Least Recently Used*) |
| L1 I-cache addressing | PIPT |
| L1 D-cache addressing | PIPT |
| L1 coherence protocol | MOESI |
| L2 cache | Shared L2 cache (1Mb) |
| L2 associativity | 16-way |
| L2 block size | 64 Bytes |
| L2 replacement policy | Pseudo random |
| L2 addressing | PIPT |
| L2 coherence protocol | MOESI |
| Interconnection | Internal CoreLink CCI-400 based |
| **Memory Subsystem** | |
| DRAM type | DDR3 x64 |
| DRAM frequency | 800MHz |
| Memory size | 2GB |
| Memory Interfaces | 1 |
| System bus frequency | 1Ghz |
| **Components** | |
| | Hard Disk, UART, 10/100 Ethernet, HDLCD and others |

Figure 30 depicts the integer pipeline stages of both Cortex-A7 and Gem5 out-of-order core model. Gem5 out-of-order model is actually based on the Alpha 21264 pipeline [GEM15f]. Note that Gem5 has a customizable pipeline, whereas the user can define the width of each stage depicted in this figure.

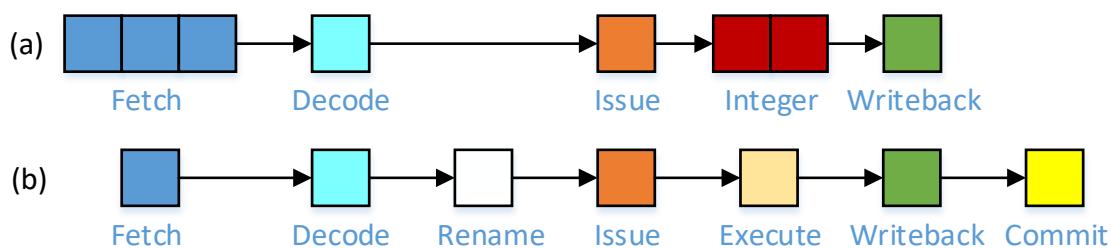

**Figure 30. Pipeline stages of (a) Cortex-A7 and (b) Gem5 out-or-order CPU model (based on [ARM11b][GEM15f]).**

---

[6] Default parameters for the Classic memory system.

Next section discusses the basic architecture mode, which will be explored in the master's degree dissertation. During the discussion, we detail the limitations found in the Gem5 framework to work with MPSoCs. In this case, our main hindrance is the simulation of multiple cores in the ARM platform.

## 6.2 Architecture Exploration

The main contribution of this work is the exploration of a diverse of MPSoC architectures based on the ARM ISA. These architectures aim to balance scalability, throughput, latency and energy consumption. The basis of this study is twofold: the separation of the memory and communication systems and the use of 3D IC to tackle the constraints described above.

We propose to disconnect the memory and communication system since they have different requirements. When a packet-based communication system is expanded to provide access to the memory system, it can be easily overburden and unable to sustain an acceptable performance [FU14][WA08][YE10]. Thus, the use of two independent system is attractive as it is more capable of maintain an acceptable performance as systems continue to scale.

For the inter-processor communication system, we propose the use of a packet-based NoC. This communication architecture, as proposed by Benini et al. [BEN02], can provide an efficient on-chip communication when compared to shared bus. Physically, distributing router units reduces the wire delays and the capacitance of the interconnection. Architecturally, decentralizing the interconnect fabric provides reliability to the system through independent operation. Figure 31(a) depicts the communication system connecting eight cores.

For the memory system we propose the use of the Gem5's CCI-like interconnect structure. This structure is heavily based on the CoreLink CCI provided by ARM. CCI aims to provide a low-cost and low-power consumption communication architecture. It is a crossbar-based communication system that provides support for cache coherence protocols between caches, I/O devices and GPUs. All read and write data channels are fixed 128-bit width. In addition, CCI supports more than one memory controller, which allows parallelism in accessing the main memory [ARM12c][ARM15g]. Figure 31(b) depicts the Gem5's CCI connecting the eight cores into a single coherent memory space. GIC (*General Interrupt Controller*) is a single controller responsible for distributing interrupts for all cores. Furthermore, since there is no core distinction for accessing the main memory this denote an UMA system. Currently, ARM does not employ NUMA in its interconnect family products [ENT15][LIN15].

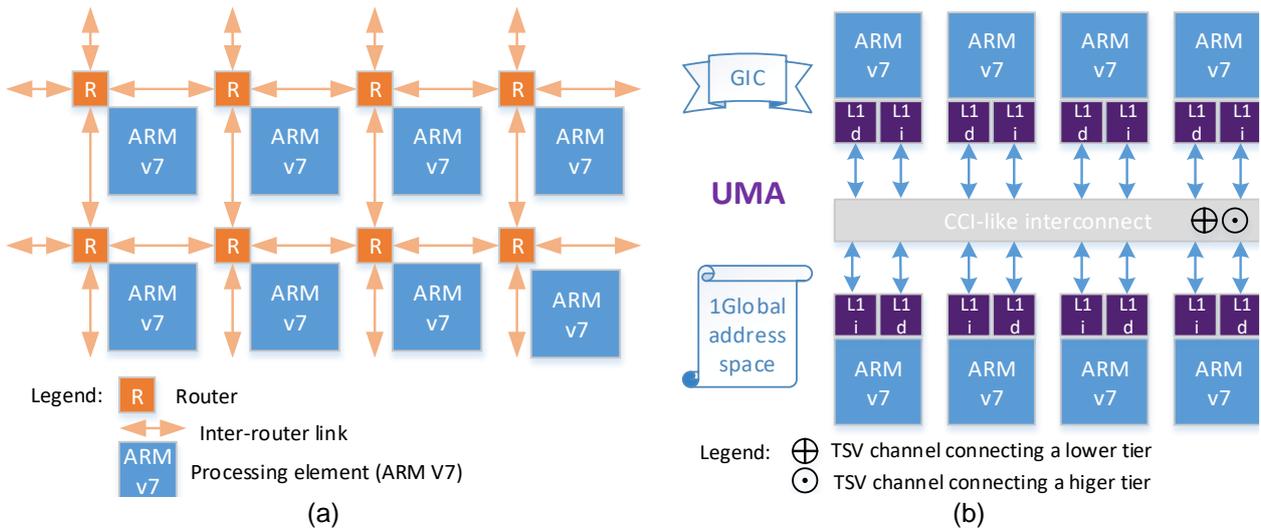

**Figure 31. (a) On-chip communication system performed with a NoC; (b) The same processors of (a) with data and code caches and the memory interconnection (Gem5's CCI).**

Figure 32 shows the entire target architecture, which comprises the junction of both, message communication and memory architectures.

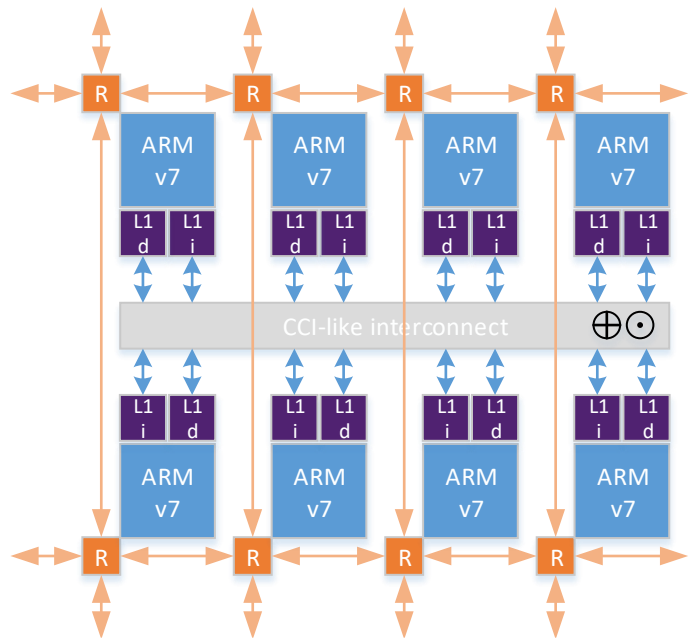

**Figure 32. A cluster of eight processors with message communication and memory architectures.**

To scale this architecture to hundreds of cores we propose to use more than one global address space, as maintaining coherence in an UMA architecture model is costly and impractical. This limitation is known as the Coherence Wall [HUA12][MAT10]. Therefore, we propose to use a hierarchy of tiers, with different models in each hierarchy level. For instance, our first approach is a two-tier system. The first tier is comprised of clusters with fully coherent UMA architecture. In this cluster, tasks are intended to be mapped according to their communication, processing and memory requirements (e.g., highly communicating task are mapped in the same cluster). The second tier is a NORMA architecture that binds multiple clusters of the first tier. In this case, tasks are intended to be mapped in different clusters that, for instance, have a sporadic communication traffic between themselves. Consequently, the one responsible for mapping tasks must take these considerations in its policy. For this work, the mapping is done by the developer – he chooses the cluster for running its application and, if he wishes so, the specific core within the cluster (through either the sched_setaffinity system call [DIE15a] or the taskset application [DIE15b]). Figure 33 depicts the memory and communication architecture in the first tier.

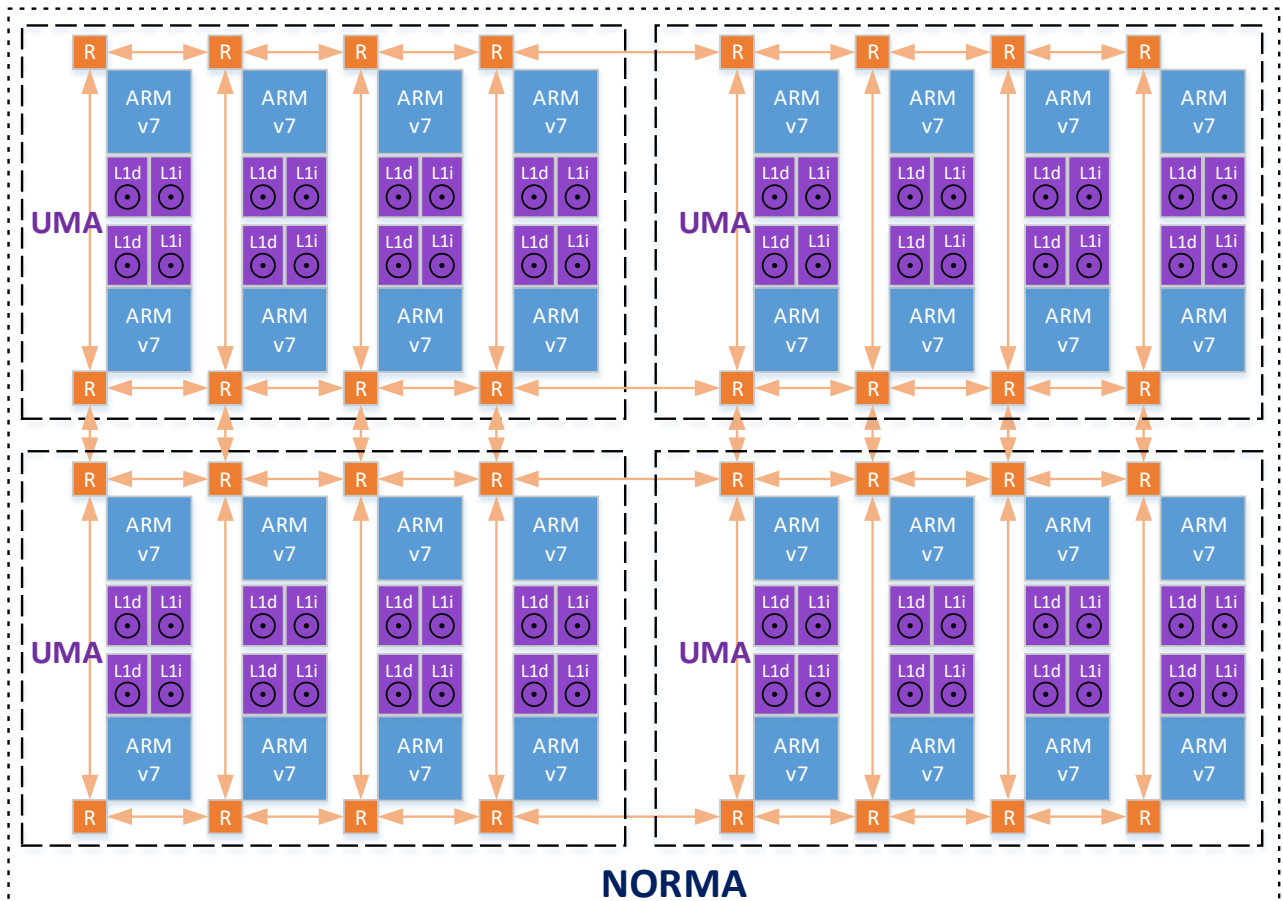

**Figure 33. First tier of the system (message communication and memory architectures).**

3D IC will be used to incorporate different levels of cache into the architecture. These levels will be shared across the number of cores attributed to each cluster. Therefore, this benefits parallel programs that share common code or data segments. In addition, we intend to incorporate emerging memory technologies into the lower levels of cache. More than one configuration will be realized to explore heat distribution and effective throughput. Figure 34 depicts an example of a two-die 3D system that is interconnected through TSVs. We will study the possibility of incorporating main memory into the chip, and, if not possible, deal with the increased number of off chip pins cause using a NORMA architecture. One possible solution is that one cluster has access to a high bandwidth memory while the others use a low bandwidth memory.

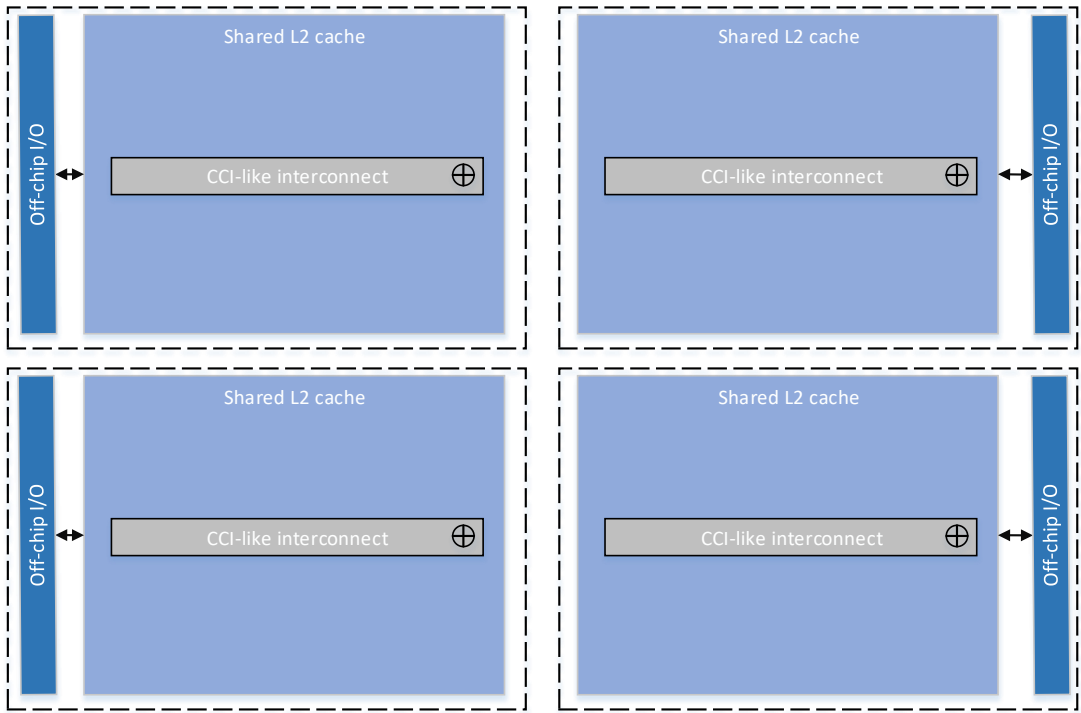

**Figure 34. Second tier of the system (Cache L2 and CCI).**

The number of tiers dimension (Z dimension) can be explored introducing new cache levels, as shown Figure 35(a) and (b), which can be connected to the tier of Figure 33.

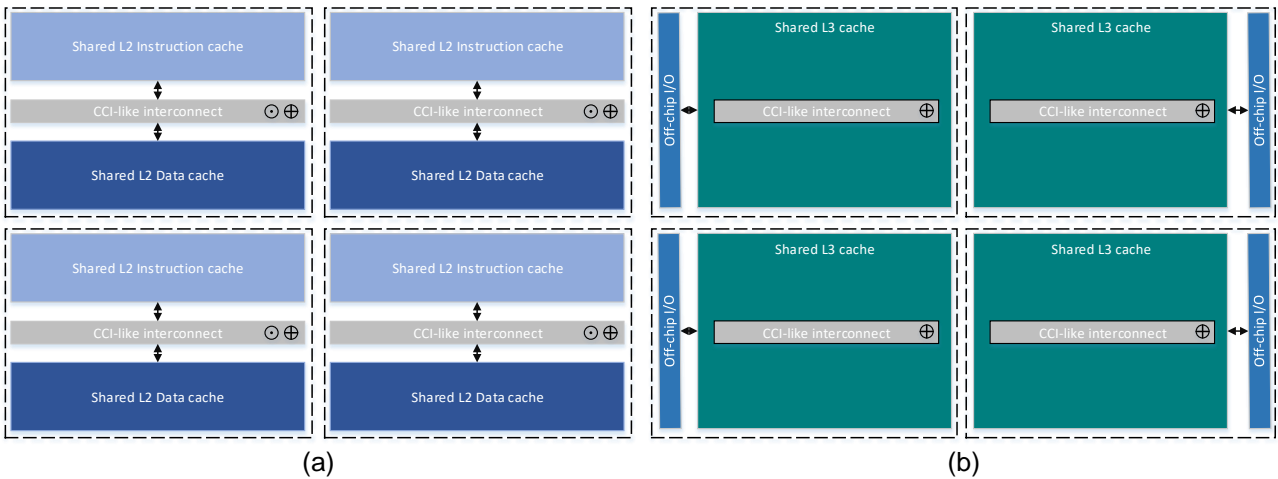

**Figure 35. (a) Level of L2 caches – instruction and data caches are separated; (b) Level of L3 caches, which mixes data and instruction.**

Additionally, the entire 3D architecture may be built to take into account several design requirements. For instance, lower levels of cache can be inserted into the inner tiers, whereas high processing elements may be inserted into the most external tires, minimizing heat problems. Figure 36 shows a possible 5-tiers MPSoC architecture. Note that, this last architecture includes UMA and NORMA programming paradigms, but also induces to a new exploration axis, since processors placed in different tires may communicate through cache hierarchy.

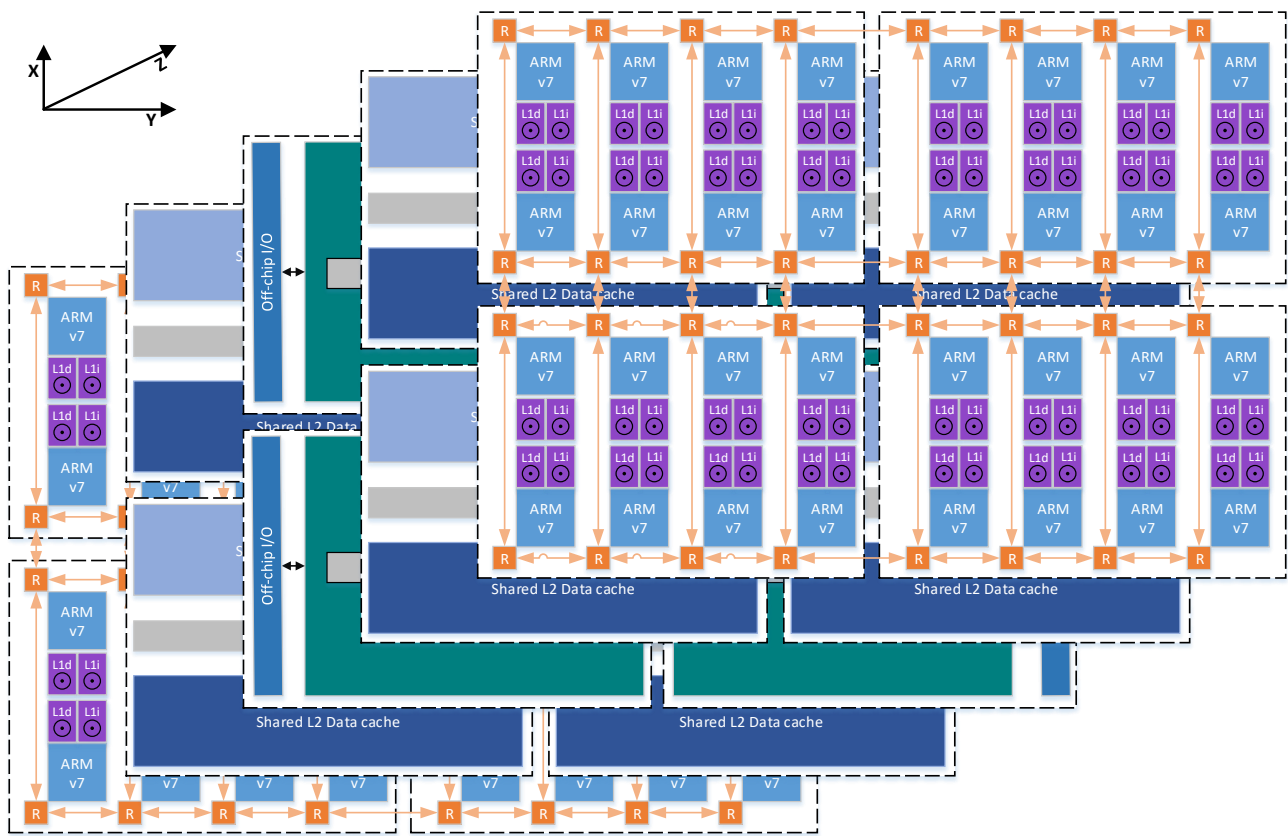

**Figure 36. Example of 3D MPSoC architecture exploration with 5 tiers. The external tiers encompasses processing elements and L1 caches. The two inner tiers connected to the external tiers includes L2 caches with separate data and code addressing. Finally, the innermost tier is a L3 cache level, which mixes data and code addressing.**

## 6.3 Limitations of the Gem5's ARM ISA

Gem5 does not impose any inherent limit of how many cores can run [GEM15g]. In spite of that, hardware and implementation nuances does limit the ability of Gem5 to simulate.

For the ARM platform, the examples provided are limited to two cores. However, systems with more than four cores present some complications [GEM15h][GEM15i]. For the Linux kernel, the platform can inform the number of cores available in many ways. For the specific version used in this work (3.3.0-rc3), the kernel consults the L2 Control Register for this. Unfortunately, this register reserves only two of its thirty-two bits to inform the cores' count [ARM15c]. This means that it is restricted to identify four core units. We expanded this field to three bits, using one bit of reserved space, expanding the previous limitation to eight cores. The pertinent code in the Linux Kernel was also modified. Note that, core identification can also be done using the SCU control register. This was not implemented, however, because there is no SCU implementation on Gem5 and this register has also the same core count limitation as the L2 register [ARM15d].

Furthermore, the GICv2 for ARM is implemented in Gem5. Regrettably, this version is restricted to service eight cores [ARM11c]. The newer version 3 of this same module can run 128+ cores. Linux kernel support was added just recently (version 4 and onward) [LWN15]. As gem5 currently stands, only eight cores can be executed in a single system.

Further difficulties were found in executing eight cores. One of the first code that every cores executes (*arch/arm/kernel/smp.c:secondary_start_kernel*) is responsible for incrementing the initial mm_struct (variable init_mm) used by the kernel. An additional increment is done on the scheduler (*kernel/sched/core.c:sched_init*). Therefore, after the system has booted, init_mm must be equal to $n + 1$, whereas $n$ is equal to the core count. If this is not met, the system will later fail with a kernel panic. Unfortunately, only the simplest model of CPU achieves this. Currently, we do not know the source of this discrepancy.

The ARM ISA on Gem5 is limited to the classic memory system. In this system, only the CCI-like interconnect is supported for on chip interconnection. This structure uses a three-layered bus for requests, responses and snoop message handling [WIE12]. Off-chip interconnection is based on this model; however, it uses a simplified two layered bus (lacks snoop messages) [GEM15k].

Internet connection is not supported on the Gem5 simulator [GEM15j]. Nevertheless, there is a way to connect multiple systems together and simulate such connection. The object to use for this is EtherLink. But, this module is limited to connect two system. Gutierrez [GUT15] extended this object to be an Ethernet Switch that connects any number of systems. However, it lacks the proper algorithm to treat congestion on an Ethernet port. Consequently, we expanded and used it in our work.

# 7 WORK SCHEDULE

3D technology is a great step in system integration, thus an intensive architectural exploration and tools are required to evaluate costs and quantify the benefits of cache hierarchy. As such, the author has already performed the following tasks:

1. **Study of related work**. Initial research on state of the art regarding cache-level exploration in 2D and 3D MPSoC architectures;

2. **Publication at 21st IEEE International Conference on Electronics Circuits and Systems (ICECS), 2014.** Publication of paper entitled "The Impact of Routing Arbitration Mechanisms on 3D NoC Latency". This work evaluates the impact of different router implementations concerning the overall latency of the system. The router implementations differ in the number of cycles needed to switch packets. This work presents latency evaluations to systems using packet based NoC as their main interconnection network, which is the case for one of the tiers of this work;

3. **Publication at 28th International Conference on VLSI Design (VLSID), 2015.** Publication of paper entitled "OcNoC: Efficient One-cycle Router Implementation for 3D Mesh Network-on-Chip". This work proposes a single cycle router implementation for 3D Mesh NoCs with two arbitration mechanisms. Results show that the proposed router can reduce latency when compared to multistage routers while maintaining minimal area overhead. This evaluation will also be considered for this work;

4. **Publication at 16th International Symposium on Quality Electronics Design (ISQED), 2015.** Publication of paper entitled "Task Partition Optimization Algorithm for Energy Saving and Load Balance on NoC-based MPSoCs". This work proposes an algorithm called Partition Reduce that is responsible for task partitioning in NoC-based MPSoCs. Experimental results were realized with two constraints: energy consumption and load balancing. Task partitioning indirectly affects the efficiency of the cache subsystem, because it can affect the hit/miss ratio of cache lines [LIN14];

5. **Presentation of PEP.** Presentation of "Projeto de Estudo e Pesquisa" required by Programa de Pós-Graduação em Ciência da Computação (PPGCC);

6. **Exploration of simulation tools for cache evaluation.** Study of available simulations tools for MPSoC with on-chip cache support. This study is presented in section 5;

7. **Architecture baseline.** To choose an architecture baseline that provides ample design space exploration for scalable and efficient 3D architectures. The architecture baseline is based on the Versatile Express board provided by ARM. Section 6 describes the details of this board.

8. **Compilation of the PARSEC benchmark suite.** The PARSEC benchmark suite was compiled for the intended ARM platform – specifically, 12 out of the 13 are ready, since the application "raytrace" is not compatible with Gem5, as it uses x86 SSE (*Streaming SIMD Extensions*) extensions for performance [GEM15I]. In addition, PARSEC provides an expanded version of the SPLASH-2 benchmark suite. They are ready as well.

Table 8 shows a summary of the complementary activities. The items below aims to detail those activities:

1. **Study of related work.** Research on state of the art regarding cache-level exploration in 3D MPSoC;

2. **Presentation of SA.** Presentation of "Seminário de Andamento" required by PPGCC;

3. **Development of Master's degree dissertation.** Development of third and final document required by PPGCC to achieve master's degree;

4. **Writing and submission of papers.** To write and to submit papers to peer-review conferences and/or journals about the architecture exploration and experimental results obtained in this work;

5. **Find a suitable message passing benchmark.** To apply and validate the architecture (more specifically, the communication system) using the packet based NoC. For this, a message passing based benchmark is necessary. If no message passing benchmark suitable for this work is found, then a synthetic application will be developed.

6. **Implementation of 3D MPSoCs with cache support.** Development of multiple architectures in the Gem5 framework as described by section 6.2;

7. **Validation of the proposed 3D MPSoCs.** To validate the accuracy of obtained results and to compare it to similar state of the art work;

8. **Evaluation of the proposed 3D MPSoCs.** To categorize and to understand results obtained; and

9. **Defense of Master's degree dissertation**.

**Table 8. Summary of Activities.**

| Activities | 2014 | | | | | | | 2015 | | |
|---|---|---|---|---|---|---|---|---|---|---|
| | 06 | 07 | 08 | 09 | 10 | 11 | 12 | 01 | 02 | 03 |
| 1. Study of related work | ■ | ■ | ■ | ■ | ■ | ■ | ■ | ■ | ■ | |
| 2. Presentation of SA | | ■ | | | | | | | | |
| 3. Master's Degree dissertation | | | | | | ■ | ■ | ■ | ■ | |
| 4. Writing and submission of paper | | | | | | ■ | ■ | | | |
| 5. Find a suitable message passing benchmark | | ■ | | | | | | | | |
| 6. Implement 3D MPSoC with cache support | ■ | ■ | ■ | | | | | | | |
| 7. Validate 3D MPSoC | | | | ■ | | | | | | |
| 8. Evaluate 3D MPSoC | | | | ■ | ■ | ■ | | | | |
| 9. Master's Degree defense | | | | | | | | | | ■ |